\documentclass[runningheads]{llncs-modified}
\usepackage{graphicx}
\usepackage{xcolor}
\usepackage{setspace}
\usepackage{amsmath, amsfonts}
\usepackage{algorithm, algorithmic}
\usepackage{array}
\usepackage{float} 
\usepackage{subfigure}
\usepackage{multirow, multicol}
\usepackage{booktabs}
\usepackage{hyperref}
\usepackage{cite}
\usepackage{comment}
\usepackage[title]{appendix}

\def\custodyscheme{$(S, \mathcal{A}, \mu)$}
\def\maxcorruptedgroupnumber{f(\gamma; S, \mathcal{A}, \mu)}

\newcommand{\E}{\mathbb{E}}

\newcommand{\symmetricdesign}{\mathcal{A}_{sym}}
\newcommand{\maxcorruptedgroupnumbersym}{f(\gamma; S, \symmetricdesign, \mu)}
\newcommand{\gammasymmetricdesign}{\gamma_{sym}}

\newcommand{\polynomialdesign}{\mathcal{A}_{poly}}
\newcommand{\gammapolynomialdesign}{\gamma_{poly}}
\newcommand{\maxcorruptedgroupnumberpoly}{f(\gamma; S, \polynomialdesign, \mu)}

\newcommand{\multilayershardingdesign}{\mathcal{A}_{mls}}

\newcommand{\randommultilayershardingdesign}{\mathcal{A}_{rmls}}
\newcommand{\maxcorruptedgroupnumberrmls}{f(\gamma; S, \randommultilayershardingdesign, \mu)}

\newcommand{\blockdesign}{\mathcal{A}_{blck}}
\newcommand{\maxcorruptedgroupnumberblck}{f(\gamma; S, \blockdesign, \mu)}
\newcommand{\gammablockdesign}{\gamma_{blck}}

\newcommand{\maxcorruptedgroupnumbernew}{f(\gamma; S, \mathcal{A}', \mu)}

\newenvironment{interposition}[1][name]{\medskip\noindent\ignorespaces\textbf{#1.}}{\medskip\noindent\ignorespacesafterend}

\hypersetup{hidelinks} 

\begin{document}
\title{Decentralized Asset Custody Scheme \texorpdfstring{\\}{}
with Security against Rational Adversary}
\titlerunning{DAC Scheme with Security against Rational Adversary}
%
\author{Zhaohua Chen\inst{1,2}
\and
Guang Yang\inst{2}
}
\authorrunning{Z. Chen, G. Yang}
%
\institute{CFCS, Computer Science Dept., Peking University \\
\and
Conflux \\
\email{chenzhaohua@pku.edu.cn} \\ \email{guang.yang@confluxnetwork.org}}
\maketitle              
\begin{abstract}
Asset custody is a core financial service in which the custodian holds in-safekeeping assets on behalf of the client. Although traditional custody service is typically endorsed by centralized authorities, decentralized custody scheme has become technically feasible since the emergence of digital assets, and furthermore, it is greatly needed by new applications such as blockchain and DeFi (Decentralized Finance).

In this work, we propose a framework of decentralized asset custody scheme that is able to support a large number of custodians and safely hold customer assets of multiple times the value of the total security deposit. The proposed custody scheme distributes custodians and assets into many custodian groups via combinatorial designs, where each group fully controls the assigned assets. Since every custodian group is small, the overhead cost is significantly reduced. The liveness is also improved because even a single alive group would be able to process transactions.

The security of this custody scheme is guaranteed under the rational adversary model, such that any adversary corrupting a bounded fraction of custodians cannot move assets more than the security deposit paid. We further analyze the security and performance of our constructions from both theoretical and experimental sides and give explicit examples with concrete numbers and figures for a better understanding of our results.

\end{abstract}


\section{Introduction}\label{sec:introduction}

Asset custody is a core financial service in which
an institution, known as the custodian, holds in-safekeeping assets such as stocks, bonds, precious metals, and currency on behalf of the client.
Custody service reduces the risk of clients losing their assets or having them stolen, 
and in many scenarios, a third-party custodian is required by regulation to avoid systematic risk.
In general, security is the most important reason why people use custody services and place their assets for safekeeping in custodian institutions.

The security of traditional asset custody service is usually endorsed by the reputation of the custodian,
together with the legal and regulatory system.
Such centralized endorsement used to be the only viable option until the emergence of blockchain and cryptocurrencies.
Cryptocurrencies enjoy two major advantages over their physical counterparts:
(1) they are intrinsically integrated with information technology such as the Internet and modern cryptography, 
which technically enables multiple custodians to safeguard assets collectively;
(2) with the underlying blockchain as a public ledger, 
the management of cryptocurrencies becomes transparent to everyone and hence any fraud behavior will be discovered immediately,
which makes prosecution much easier. 

From a systematic point of view,
asset custody service provided by a federation of multiple independent custodians has better robustness and resistance against single-point failure,
and hence achieves a higher level of security.
Such credit enhancement is especially important for the safekeeping of cryptoassets on decentralized blockchains such as Bitcoin~\cite{Nakamoto08} and Ethereum~\cite{Wood14}, 
where the legal and regulatory system is absent or at least way behind the development of applications.
For example,
in the year 2019 alone, at least $12$ cryptocurrency exchanges claimed being hacked and loss of cryptoassets totaled to around $2.9$ billion dollars~\cite{web_hacked}.
However, it is difficult for customers to distinguish that whether the claimed loss was caused by a hacker attack or internal fraud and embezzlement,
and therefore raises the need for decentralized asset custody.

Decentralized asset custody finds applications in many scenarios related to blockchain and digital finance. 
A motivating example is the \emph{cross-chain assets mapping} service (a.k.a. \emph{cross-chain portable assets}~\cite{ZamyatinAZKMKK19,buterin2016chain}) which maps cryptoassets on one blockchain to tokens on another blockchain for inter-chain operability. 
For instance, the mapping from Bitcoin to Ethereum
enables usage of tokens representing bitcoins within Ethereum ecosystem, 
and in the meanwhile, the original bitcoins must be safeguarded so that the bitcoin tokens are guaranteed redeemable for real bitcoins in full on the Bitcoin network.
Nowadays the volume of cryptoassets invested into Ethereum DeFi applications 
is massive, and the highest point in history almost reaches 90 billion dollars~\cite{web_defi},
among which a significant fraction (e.g. H-Tokens~\cite{H-Tokens}, imBTC~\cite{imBTC}, tBTC~\cite{tBTC}, WBTC~\cite{WBTC}, renBTC~\cite{renBTC}, etc.) is mapped from Bitcoin.
Due to the reality that most of those DeFi applications and tokens remain in a gray area of regulation,
decentralized cryptoassets custody turns out an attractive approach for better security and credit enhancement.

In this work, we propose a framework of decentralized asset custody scheme designed for cross-chain assets mapping (especially from blockchains with poor programmability, e.g. Bitcoin).
More specifically, custodians and assets are distributed into multiple custodian groups, 
where each group consists of few custodians as its members and fully controls a small portion of all assets under custody.
The authentication of each custodian group requires the consent of sufficiently many group members, 
which can be implemented with voting or threshold signature.
Under this framework, transactions can be processed more efficiently within the very few group members,
since the computational and communicational cost is significantly reduced.
The liveness and robustness are also improved since even a single alive custodian group can process transactions.

The security of our proposed asset custody scheme is guaranteed against a rational adversary:
every custodian in this scheme must offer a fund as the security deposit,
which is kept together with the asset under custody and will be used to compensate for any loss caused by misbehaving custodians. 
The system remains secure as long as an adversary cannot steal more assets than the deposit paid,
i.e. comparing to launching an attack the adversary would be better off by just withdrawing the security deposit of custodians controlled.
Furthermore, we prove that for an adversary who corrupts a limited fraction of custodians, 
our scheme can safeguard customer assets of multiple times the value of the total security deposit under suitable construction.
This approach significantly reduces the financing cost of a collateralized custody service.

\subsection{Related Works}\label{sec:intro-related-works}

The prototype of decentralized asset custody scheme first appears in Bitcoin as multisignature (multisig) \cite{web_multisig},
where the authentication requires signatures from multiple private keys rather than a single signature from one key.
For example, an $M$-of-$N$ address requires signatures by $M$ out of totally $N$ predetermined private keys to move the money.
This na\"{i}ve scheme works well for small $M$ and $N$ but can hardly scale out,
because the computational and communicational cost of authenticating and validating each transaction grows linearly in $M$. 
Both efficiency and liveness of the scheme are compromised for large $M$ and $N$, especially in the sleepy model proposed by Pass and Shi~\cite{PS17} where key holders do not always respond in time.
In practice, a multisignature scheme is typically used at the wallet level rather than as a public service,
since the scheme becomes costly for large $N$ and most Bitcoin wallets only support $N\le 7$. We remark that
multisignature schemes may be coupled with advanced digital signature techniques such as threshold signature~\cite{GagolKSS20,BonehGG17} or aggregate signature~\cite{schnorr1991efficient,bellare2006multi,maxwell2019simple}
to reduce the cost of verifying multi-signed signatures.

As for the cross-chain asset mapping service,
existing solutions mainly include the following types:
\begin{itemize}
	\item Centralized: custody in a trusted central authority, with the endorsement fully from that authority, e.g.
	H-Tokens~\cite{H-Tokens}, WBTC~\cite{WBTC} and imBTC~\cite{imBTC}; 
	
	\item Consortium: custody in multisignature accounts controlled by an alliance of members, and endorsed by the reputation of alliance members, e.g. cBTC~\cite{cBTC} (in its current version) and Polkadot~\cite{Wood16};
	
	\item Decentralized (with deposit/collateral): 
	custody provided by permissionless custodians,
	with security guaranteed by over-collateralized cryptoassets,
	e.g. tBTC~\cite{tBTC} and renBTC~\cite{renBTC} (in its future plan).
\end{itemize}

The last type seems satisfiable in decentralization and security against single-point failure and collusion. Meanwhile, existing solutions (tBTC and renBTC) have security guaranteed in the sense that an adversary will not launch a non-profitable attack.
However, for these solutions, significant drawbacks exist as well. The first drawback is the inefficiency caused by over-collateralization,
e.g. tBTC requires the custodian to provide collateral worth of $150\%$ value of customer's assets,
and renBTC requires $300\%$.
The second drawback is that 
these solutions cannot support homogeneous collateral as the assets under custody.
Otherwise, an adversary corrupting a single group would be able to steal more than the collateral paid, therefore breaking the safety of the custody service in market volatility. 
We remark that~\cite{HarzGGK19} considers the dynamic adjustment of the deposit of custodians in the long run.
However, this work implicitly assumes that the security of the system is irrelevant with the behavior of custodians (e.g. by introducing cryptographic methods like in Bitcoin). Such assumption is inapplicable in the game-theoretic setting we discuss here.


\subsection{Our Contributions}\label{sec:intro-contribution}

Our contributions lie in the following six parts:
\begin{itemize}
	\item In literature, we are the first to consider the possibility of homogeneously keeping exterior assets and custodians' deposit in the scenario of decentralized asset custody. To model such feasibility, we formalize the concept of custody scheme and further propose the concept of efficiency factor of a custody scheme for any adversary power (Section~\ref{sec:model}). The latter captures the maximal ratio of capable exterior assets to deposit that the underlying custody scheme can safely handle against a rational adversary. 
	\item We propose a series of evaluation criteria to specify the performance of a custody scheme (Section~\ref{sec:model}). Combining with the previous point, we give a complete framework for analyzing a custody scheme and comparing different custody schemes. We point out that the underlying group assignment scheme is the core of a custody scheme.
	\item We present four kinds of concrete construction of group assignment schemes. For each of them, we theoretically give an exact value/a lower bound on the efficiency factor of the custody scheme they induce (Section~\ref{sec:construction}, Appendix~\ref{app:multi-layer-sharding-design}). Some results turn out to be magnificent. For example, we show that we can assign $24$ custodians to $759$ groups such that as long as the adversary corrupts $\gamma\le 1/4$ fraction of all custodians,	the custody scheme is capable of safekeeping assets worthy of $\eta>30.62$ times of total collateral.
	\item We prove that the random sampling trick significantly reduces the size of group assignment scheme without losing too much in the efficiency factor (Section~\ref{sec:random-sampling}). Therefore, random sampling resolves the problem of too many groups inside a custody scheme.
	More specifically, suppose we have a custody scheme consisting of $n$ participants and its efficiency factor is $\eta$ against some adversary.	By randomly sampling ${O}\left( \eta n \right)$ many groups, the newly induced custody scheme would have efficiency factor $\eta'\ge \sqrt{\eta + 1}-2$ against the same adversary with high probability. An important corollary shows that we can construct $\Theta(n)$ groups with identical size $\Theta(1)$ to obtain an efficiency factor of $\Theta(1)$ against an adversary with constant power.
	\item For the complexity issue, we prove that it is NP-hard to find an optimal corrupting strategy in general. However, given a group assignment, we show that within polynomial time, we can find a solution no worse than the average case (Appendix~\ref{app:hardness}). 
	\item We conduct extensive experiments to reveal the real-life performance of the group assignment scheme designs we propose (Appendix~\ref{app:experimental-results}), also as a complement to our theoretical study. We further compare these designs according to the evaluation criteria we present. As an accessory, we expose the potential positive correlation between the efficiency factor and the number of custodian groups. Nevertheless, complicated assignments with too many groups may be infeasible to manage in practice. 
\end{itemize}

For writing smoothness, proofs of all propositions and theorems in the main body are deferred to Appendix~\ref{app:proofs}.


\section{Model}\label{sec:model}

Our goal is to implement the decentralized custody scheme without relying on any trusted party. 
More specifically, we investigate the feasibility that $n$ custodians (a.k.a. $n$ \emph{nodes})
jointly provide the custody service, 
such that the security is guaranteed as long as 
a bounded fraction of custodians are corrupted, e.g. no more than $n/3$ nodes are corrupted simultaneously. 
This assumption of an honest majority is much milder than assuming a single party trusted by everyone, and hence likely leads to a better security guarantee in practice.

The decentralized custody scheme is based on overlapping group assignments.
That is, custodians are assigned to overlapping groups, and each group is fully controlled by its members
and holds a fraction of the total assets under custody, including both deposit from custodians and assets from customers.
In what follows we assume that the in-safekeeping assets are \emph{evenly} distributed to custodian groups,
since an uneven distribution naturally leads to degradation of security and capital efficiency.

Furthermore, we consider the security of a custody scheme against a \emph{rational adversary}:
the adversary may corrupt multiple nodes, but will not launch an attack if the potential profit does not exceed the cost.
To achieve security under such a model, 
every custodian in our scheme must provide an equal amount of deposit, which will be confiscated and used for compensation in case of misbehavior. 
Thus, 
if misbehavior can be detected in time,
no rational adversary would ever launch an attack as long as the deposit paid outweighs the revenue of a successful attack. 
Here, we emphasize that instead of resorting to another level of collateral custody service, the deposit from custodians is maintained as a part of the total assets under custody, together with assets from external customers.

As a remark, we assume that attacks in the decentralized custody scheme can be detected immediately.
If the decentralized custody service is for cryptoassets and deployed on a blockchain,
then all instructions from customers and transfer of assets are transparent to everyone,
and hence any malicious transaction will be caught immediately.
Alternatively, the detection may be implemented with the periodic examination which ensures that misbehavior is discovered before the adversary can exit or change the set of corrupted nodes.
In other scenarios, detecting corrupted behavior may be a non-trivial problem, but for the sake of this study we will leave it out to avoid another layer of complication.

The incentive of agents participating in this collateralized custody scheme is also indispensable for a full-fledged decentralized custody service.
A reasonable rate of the commission fee and/or inflation tax would be sufficient to compensate the cost of agents providing such custody service.
In the blockchain scenario, an extra per-transaction fee is also an option.
Overall we believe that the mechanism design to incentivize custodians is essentially another topic,
which is beyond the scope of this work and should be left for future study.

\begin{interposition}[A trivial but useless solution]
	In the most trivial solution, the asset under custody can only be moved when approved by all custodians or at least a majority of them.
	However, as $n$ grows getting such an approval becomes expensive and even infeasible in practice,
	especially when honest participants may go off-line (as in the sleepy model~\cite{PS17}),
	which renders the trivial scheme useless. 
\end{interposition}

Although the above solution is not satisfactory, 
it does provide enlightening ideas for designing a better custody scheme.
The threshold authorization scheme guarantees that the adversary cannot move any assets under custody if not a sufficient number of nodes are corrupted.
More generally, this is a specific case of security against the rational adversary, where with bounded power,
the adversary's deposit outweighs the revenue of launching an attack.
Again, as long as this property is satisfied the custody scheme is secure in our model.

In particular, the following toy example shows the feasibility of implementing our idea with multiple overlapping subsets of $S$ as custodian groups.
In this example, \emph{each} 3-subset of $S$ controls a certain fraction of the total assets under custody. Here $S$ is the set of all custodians.

\begin{example}[Toy example]\label{exa:toy}
    Consider the case when $10$ units of exterior assets are under custody. Assume there are $n = 5$ custodians, each paying a deposit of $6$ units of assets, amounting to $30$ units.
    Let each of the $10$ 3-subsets of $S$ form a custodian group,
    and assign all $40$ units of assets equally to all groups, i.e. each custodian group controls $4$ units.
    If the asset controlled by each group can be moved with approval of $2$ out of $3$ members in that group,
    then an adversary controlling $2$ nodes can corrupt exactly $3$ custodian groups.
    However, by controlling $3$ groups the adversary can only move $4\times 3 = 12$ units, which is no more than the deposit of corrupted nodes (also $12$ units).
    Thus such a custody scheme for $n = 5$ is secure against adversaries controlling up to two nodes.
\end{example}

In what follows, we will formalize the model of a decentralized custody scheme with assets evenly distributed among custodian groups. 
To start with, we introduce a formal definition of the custody scheme we consider in this work.

\begin{definition}[Custody scheme]\label{def:custody-scheme}
	A custody scheme \custodyscheme\ consists of the following three parts: 
    \begin{itemize}
        \item $S = \{1, 2, \cdots, n\}$ denotes the set of all custodians (or simply nodes);
        \item $\mathcal{A}$ denotes a family of $m$ $k$-subsets of $S$,
        such that each element in $\mathcal{A}$ (i.e. a $k$-subset of $S$) represents a custodian group under the given custody scheme;
        \item $\mu \in [1/2, 1)$ denotes a universal authentication threshold for all custodian groups,
        i.e. the asset controlled by that group can be settled arbitrarily with approval of \emph{strictly above} $\mu k$ group members.
    \end{itemize}
\end{definition}

We emphasize that the elements in $\mathcal{A}$ do not have to be disjoint.
In fact, it is imperative to use overlapping subsets in any meaningful solution. In certain cases, there might even exist repeated elements in $\mathcal{A}$.

In this work, we focus on the symmetric setting where every node provides the same amount of deposit and every custodian group has the same fraction of total assets in custody.
At the same time, our discussion of the authentication threshold $\mu$ mainly focuses on $\mu = 1/2$ and $\mu = 2/3$.
\footnote{In a synchronous network, $\mu \geq 1/2$ is a sufficient condition for the existence of expected-constant-round Byzantine agreement protocols in the authenticated setting (i.e., with digital signature and public-key infrastructure)~\cite{KatzK09}, whereas $\mu \geq 2/3$ is necessary and sufficient for the existence of Byzantine agreement protocols in the unauthenticated setting~\cite{PeaseSL80}.
We further remark that smaller $\mu$ implies less security but better liveness, for example, when $\mu \to 0$, even a single corrupted member can block a custodian group.
However, the discussion of liveness is beyond the scope of this work.} 
We let $r=\lceil \mu k + \epsilon \rceil$ denote the smallest integer greater than $\mu k$, 
and hence the authentication of every custodian group is essentially an $r$-of-$k$ threshold signature scheme.

We represent the adversary power with $\gamma \in (0, 1)$,
which refers to the fraction of corrupted nodes in $S$. 
Specifically, we let
$s=\lfloor \gamma n \rfloor$ denote the number of corrupted nodes in $S$.
\footnote{In most parts of the paper, we slightly abuse the notation and assume that $\gamma n$ is always a natural number, i.e. $s=\gamma n\in \mathbb{N}$.}
The adversary is allowed to adaptively select corrupted nodes and then get all information and full control of those nodes thereafter, as long as the number of corrupted nodes does not exceed $s$.
In case a group in $\mathcal{A}$ contains \emph{at least} $r$ corrupted nodes,
we say that group is \emph{corrupted}.
Furthermore, we remark that the adversary has reasonably bounded computing power, 
so that cryptographic primitives such as digital signatures are not broken.

Given a custody scheme $(S,\mathcal{A},\mu)$, together with $\gamma$ for the adversary power,
we use the function $\maxcorruptedgroupnumber$ to denote the maximal number of groups that may be corrupted. Formally,
\begin{equation}\label{def:maxcorruptedgroupnum}
	\maxcorruptedgroupnumber := \max_{B\subseteq S: |B|= \lceil\gamma n\rceil}
	\left|\{ A\in \mathcal{A} \mid |A \cap B|> \mu k \}\right|.
\end{equation}

Recall that as all assets under custody are equally distributed to all custodian groups, 
each corrupted group values equal to the adversary. Therefore, $\maxcorruptedgroupnumber$ directly resembles the maximal gain of the adversary.

We further define the efficiency factor of a custody scheme, which captures the ability to securely holding exterior assets.
\begin{definition}[Efficiency factor of a custody scheme]\label{def:efficiency-factor}
	Given a custody scheme $(S, \mathcal{A},\mu)$ and adversary power $\gamma$ defined as above, 
	the \emph{efficiency factor} of this scheme against $\gamma$-adversary, denoted by $\eta$, 
	is defined as:
	\[\eta := \frac{\gamma\cdot m}{\maxcorruptedgroupnumber} - 1.\]
	where $m$ is the total number of custodian groups induced by $\mathcal{A}$.
\end{definition}

The efficiency factor $\eta$ indeed equals the maximal ratio of capable exterior assets to deposit that the underlying custody scheme can handle. Specifically, 
suppose that $u$ units of assets are deposited in total, and $v$ units of exterior assets are in custody. 
According to~(\ref{def:maxcorruptedgroupnum}), by launching an attack the adversary is able to seize the funds of $\maxcorruptedgroupnumber$ custodian groups, 
which amounts to
$(u + v) \cdot {\maxcorruptedgroupnumber} /{m}$ units of assets,
at the cost of losing deposit worthy of value $\gamma\cdot u$ units. 
Recall that in our model, collateral and exterior assets are homogeneous and kept together by the custodian groups, therefore, the custody scheme is secure as long as $\maxcorruptedgroupnumber/m\cdot (u + v) \leq \gamma\cdot u$,
or equivalently,
$v/u\leq \eta$ according to Definition~\ref{def:efficiency-factor}.

As an example for the definition, $\eta = 1$ implies that the system is secure when the total value of exterior assets is no more than the total value of deposit.

Notice that when the efficiency factor $\eta < 0$ for some $\gamma$, the custody scheme against that $\gamma$-adversary is always insecure, 
regardless of the amount of deposit. 
To capture such property, we further define the reliability and safety of a custody scheme based on the Definition~\ref{def:efficiency-factor}.

\begin{definition}[Reliability and safety of custody scheme]\label{def:reliability-security}
	For a custody scheme $(S, \mathcal{A}, \mu)$ and adversary power $\gamma$, 
	we say that the custody scheme is \emph{$\gamma$-reliable} if the efficiency factor $\eta$ of the scheme is non-negative against $\gamma$-adversary, 
	i.e. $\maxcorruptedgroupnumber \le \gamma \cdot m$.
	Furthermore, the scheme is \emph{secure against $\gamma$-adversary} (or simply \emph{secure}) 
	if it is $\gamma'$-reliable for every $\gamma'\in(0, \gamma]$.
\end{definition}

Putting into our formal definition, 
the trivial solution with only one custodian group (i.e. $k=n$, $m=1$) has efficiency factor $\eta=\infty$ for $\gamma \leq \mu$ and $\eta<0$ for $\gamma > \mu$; 
the custody scheme in Example~\ref{exa:toy} has its efficiency factor $\eta$ changing
according to the adversary power $\gamma$ as summarized in Table~\ref{tab:toy}. In particular, for $\gamma = 1/5$ and $\gamma = 2/5$, the scheme is reliable with $\eta=\infty$ and $\eta = 1/3$ respectively.
For $\gamma \ge 3/5$ the scheme is unreliable with $\eta < 0$.

\begin{table}[!htbp]
	\caption{The efficiency factor of the custody scheme under different adversary power in Example~\ref{tab:toy}.}
	\label{tab:toy}
	\centering
	\renewcommand{\arraystretch}{1.35}
	\setlength{\tabcolsep}{5pt}
	\begin{tabular}{l*{4}{c}}	
		\toprule
		{parameters $\backslash$ adversary power ($\gamma$)} & $1/5$ & $2/5$ & $3/5$ & $4/5$ \\
		\midrule
		\# corrupted nodes ($s$) & $1$ & $2$ & $3$ & $4$  \\ 
		\# corrupted custodian groups ($\maxcorruptedgroupnumber$) & $0$ & $3$ & $7$ & $10$ \\
		efficiency factor ($\eta$) & $\infty$ & $1/3$ & $-1/7$ & $-1/5$ \\
		\bottomrule
	\end{tabular}
	\smallskip\\
	\flushleft\textit{
		The authentication threshold is realized as $r=2$ and $\mu=1/2$ (in this example equivalent to have $\mu\in [1/3,2/3)$).
	}
\end{table}

From the formalization of our decentralized custody scheme,
it is clear that the custodian group assignment $\mathcal{A}$ is the core of the whole custody scheme.
In particular, for a fixed $n$, every specific group assignment $\mathcal{A}$ and fixed constant $\mu$ (say, $\mu\in\{1/2, 2/3\}$),
as the parameters $m$ and $k$ are already specified in $\mathcal{A}$,
the maximal number of corrupted groups and
the efficiency factor $\eta$ are functions solely depending on the adversary power $\gamma$.\footnote{We remark that the number of custodians $n$ is not always extractable from the group assignment scheme $\mathcal{A}$, as in some cases, especially when we consider random sampling in Section~\ref{sec:random-sampling}), some custodians may belong to no group.
}

Therefore, in the rest of this paper, we will focus on the construction and analysis of custodian group assignment schemes.
In the meantime, we point out that it is meaningless merely to study a single group assignment scheme. Even in real life, the group assignment scheme should be adjusted with the joining and leaving of custodians. Instead, we focus on the systematic construction methods which lead to group assignment scheme \emph{families}. 
\begin{definition}[Group assignment scheme family]
	We say $\mathcal{C} = \{\mathcal{A}^n\}_{n\in \mathcal{I}}$ is a group assignment scheme family, if
	\begin{itemize}
		\item $\mathcal{I}$ is an index set;
		\item $\mathcal{A}^n$ is a group assignment scheme with $n$ nodes;
		\item all group assignment schemes in $\mathcal{C}$ imply an identical group size.
	\end{itemize} 
\end{definition}	

\begin{interposition}[Evaluation criteria]
	In this work, we use the following evaluation criteria when comparing two group assignment scheme families with the same group size:
	\begin{enumerate}
		\item \textit{Efficiency factor.} Firstly, we consider the efficiency factor $\eta$ of schemes in two families with the same number of nodes under adversary power $\gamma = 1/2\cdot\mu, 2/3\cdot\mu$. We prefer the family with a higher efficiency factor of group assignment schemes.
		\item \textit{Number of groups.} Secondly, we consider the size $m$ of schemes in two families with the same number of nodes. We prefer the family with less size of group assignment schemes. In real life, a large amount of groups leads to a high maintenance cost of the custody scheme.
	\end{enumerate}
\end{interposition}

\section{Constructions of Group Assignment Schemes}\label{sec:construction}

In this section, we propose three types of group assignment schemes and analyze the performance of resultant custody schemes. 
We also provide empirical analysis of these schemes with concrete numbers for a better understanding. We leave another type of group assignment scheme to Appendix~\ref{app:multi-layer-sharding-design}.

\subsection{Symmetric Design}\label{sec:symmetric-design}

\begin{construction}[Symmetric design]\label{constr:symmetric-design}
    Given $n$ and $k$, let $\symmetricdesign$ be a family consisting of all $k$-subsets of $S$ as custodian groups,
    i.e. $\symmetricdesign$ is an assignment with $m = \binom{n}{k}$ different groups where each group has $k$ nodes.
    For every authentication threshold $\mu$, 
    a custody scheme is induced by $\symmetricdesign$ and $\mu$.
\end{construction}

Due to the perfect symmetry of $\symmetricdesign$,
it immediately follows that the number of corrupted groups in the above custody scheme only depends on the number of corrupted nodes. 
For the adversary corrupts any set of $\gamma n$ nodes, the number of corrupted groups can be calculated as follows:
\begin{equation}
    \maxcorruptedgroupnumbersym =
    \sum_{r\le t\le k} \binom{\gamma n}{t}\binom{n-\gamma n}{k-t}.
 \end{equation} 

%
The efficiency factor turns out to be $\eta =
\gamma\cdot \left.\binom{n}{k}\middle/\sum_{t = r}^{k}\binom{\gamma n}{t}\binom{n - \gamma n}{k - t}\right. - 1.$ When $\mu \geq \gamma$,\footnote{We mention that in this work, when considering the reliability of a custody scheme, we tacitly approve that $\mu \geq \gamma$. For a better understanding, consider the first example with only one group consisting of all custodians. Under such group assignment, when $\gamma > \mu$, the scheme is surely $\gamma$-unreliable.} according to the tail bound of hypergeometric distribution~\cite{chvatal79}, we have
\begin{align}
	\eta = \gamma\cdot \left.\binom{n}{k}\middle/\sum_{t = r}^{k}\binom{\gamma n}{t}\binom{n - \gamma n}{k - t}\right. - 1 \geq \gamma\cdot e^{2(\gamma - \mu)^2k} - 1, \label{ieq:symmetric-design}
\end{align}
which establishes a good lower bound on the efficiency factor of the symmetric design under appropriate $\gamma$.

In the following proposition, we demonstrate that for appropriately large $k$, $\symmetricdesign$ is secure for $\gamma$ close to $\mu$. The proof is provided in Appendix~\ref{app:proof-prop-symmetric-design-1}.

\begin{proposition}\label{prop:symmetric-design-1}
	For any $k$ and $n$, given $\mu$ and corresponding $r = \lceil \mu k + \epsilon \rceil$, if $\sqrt{2(r - 1)\ln \frac{k - 1}{r - 1}} < \min\{r - 1, k - r\}$, then the custody scheme induced by $\symmetricdesign$ and $\mu$ is secure against $\gammasymmetricdesign$-adversary, for $\gammasymmetricdesign$ defined as follows:
	\[\gammasymmetricdesign := \frac{r - 1 - \sqrt{2(r - 1)\ln \frac{k - 1}{r - 1}}}{k - 1}.\]
\end{proposition}

For the special case when $n$ is even, $k$ is odd, $n \geq 2k$ and $\mu = 1/2$, the security threshold of custody scheme induced by symmetric design can be enhanced to $1/2$, as shown in the following proposition. The proof of the proposition is deferred to Appendix~\ref{app:proof-prop-symmetric-design-2}.

\begin{proposition}\label{prop:symmetric-design-2}
	For any odd $k$ and even $n$ with $n \geq 2k$, the custody scheme derived from $\symmetricdesign$ and $\mu = 1/2$ is secure against $1/2$-adversary.
\end{proposition}

\begin{figure}[!t]
    \centering
	\includegraphics[width=\textwidth]{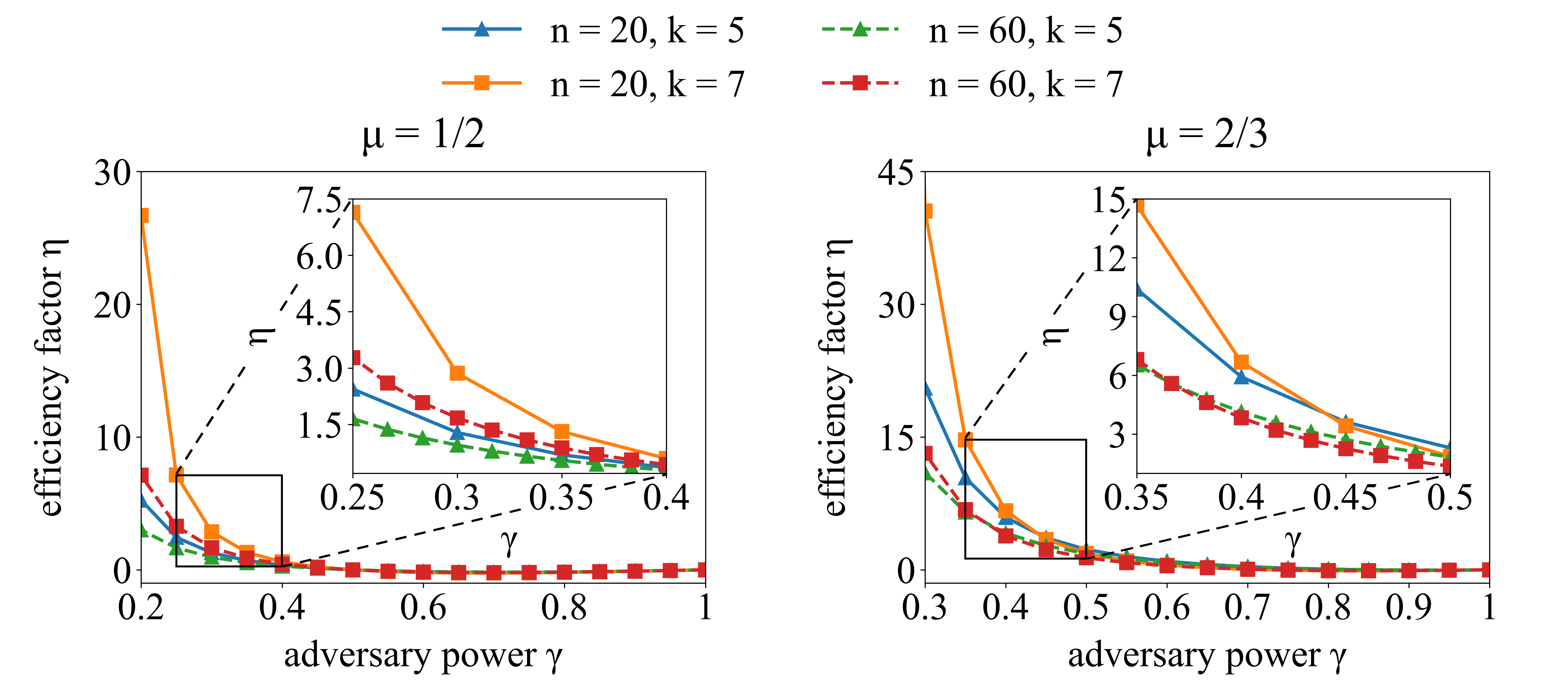}
    \caption{The efficiency factor $\eta$ against adversary power $\gamma$ for $\symmetricdesign$ as in Construction~\ref{constr:symmetric-design}. In particular, $\eta < 0$ iff the custody scheme is not secure for the corresponding $\gamma$.}
    \label{fig:symmetric-design}
\end{figure}

Fig.~\ref{fig:symmetric-design} depicts the relation between efficiency factor $\eta$ and adversary power $\gamma$,
for $n\in\{20,60\}$, $k\in\{5,7\}$, and $\mu\in\{1/2, 2/3\}$.
Basically, we see that with fixed $n$, $k$ and $\mu$, the efficiency factor $\eta$ of the custody scheme induced by symmetric design decreases as $\gamma$ grows. Further, for combinations of reasonably large $n$ and $k$, 
the efficiency factor $\eta$ can be above $10$ when $\gamma$ is roughly $1/2\cdot \mu$.
For instance,
when $n = 20$, $k = 5$ and $\mu = 2/3$, 
we have $m = \binom{20}{5} = 15,504$ and the efficiency factor $\eta = 10.4$ against adversary with power $\gamma = 0.35$.

\begin{figure}[!t]
    \centering
	\includegraphics[width=\textwidth]{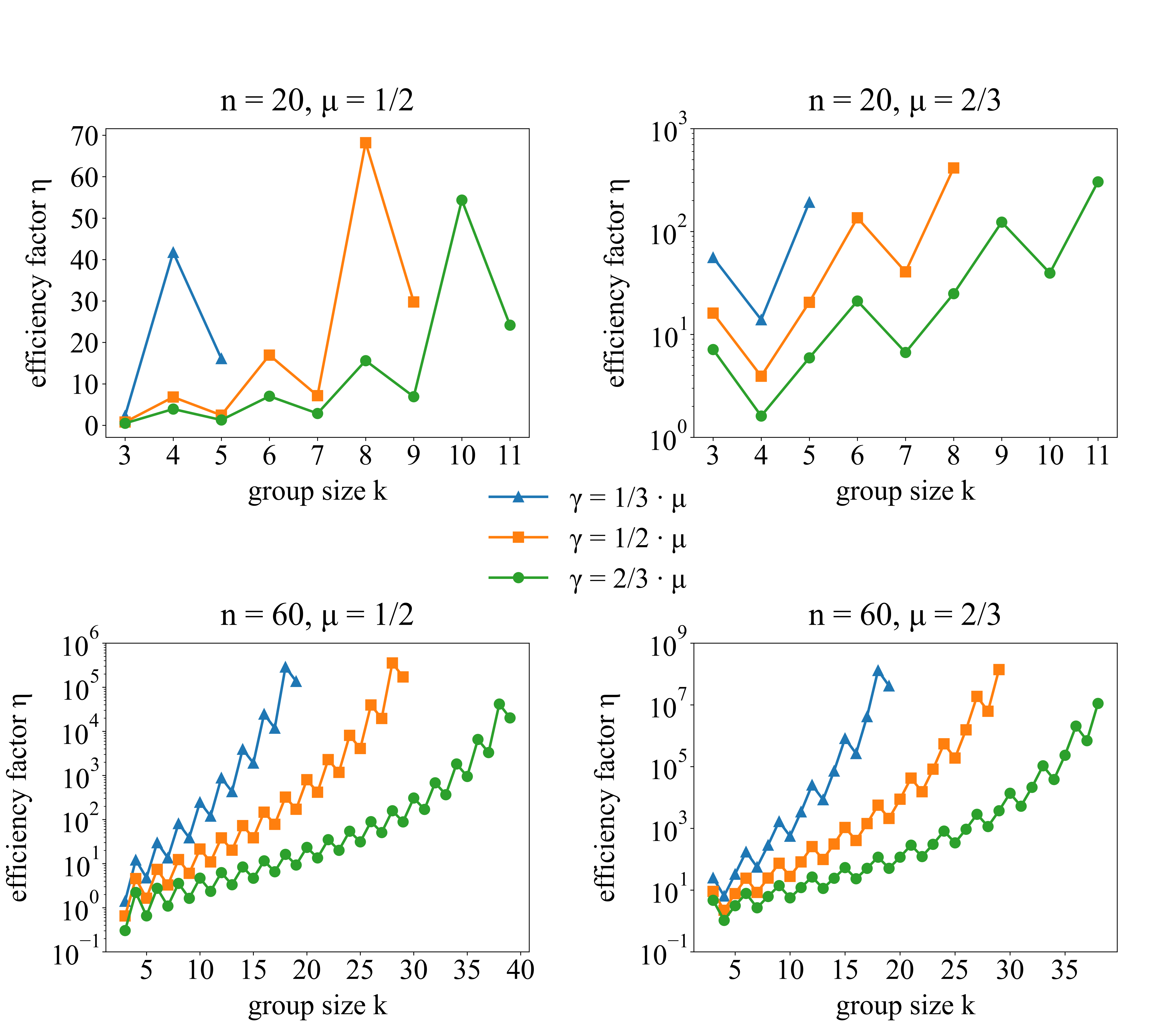}

    \caption{The efficiency factor $\eta$ against group size $k$ for $\symmetricdesign$ as in Construction~\ref{constr:symmetric-design}. Blank points on the right side refer to $\eta=\infty$ when adversary cannot corrupt even a single custodian group.}
    \label{fig:symmetric-design-group-size}
\end{figure}

Fig.~\ref{fig:symmetric-design-group-size} illustrates the behavior of the efficiency factor $\eta$ versus the custodian group size $k$, for $n \in\{20, 60\}$, $\mu \in\{1/2, 2/3\}$ and $\gamma \in\{1/3\cdot \mu, 1/2\cdot \mu, 2/3\cdot \mu\}$.
The figure shows that in general, $\eta$ increases with $k$ for custody schemes induced by $\symmetricdesign$.
The sawteeth appearing on the curves are due to the rounding of $r$ and $s$,
i.e. the authentication threshold and the number of corrupted nodes.

Finally we remark that the construction of $\symmetricdesign$ by itself is mainly a theoretical result. 
Because the size of such group assignment $m = \binom{n}{k}$ grows too fast and hence $n$ and $k$ must be severely bounded in practice, 
e.g. $n \sim 20$ and $k \sim 5$,
in order to keep $m$ reasonable.
One solution to mitigate the above issues is by random sampling,  
as exhibited in Section~\ref{sec:random-sampling}.

\subsection{Polynomial Design}\label{sec:polynomial-design}

The following construction of group assignments 
relies on polynomial-based combinatorial designs.

\begin{construction}[Polynomial design]\label{constr:polynomial-design}
	For given $k$, let $q\geq k$ be a prime and the number of custodians be $n = kq$. Let $T = \{(a, b) \;|\; 0\leq a\leq k - 1, 0\leq b \leq q - 1\}$ be a set of size $kq$, therefore, there is a bijection from $S$ to $T$. (For simplicity, we use an element in $T$ to represent the unique corresponding element $S$.) At last, let $0 < d < k$ be a integer. The polynomial design $\polynomialdesign$ is a family of $m = q^{d}$ $k$-subsets of $S$ defined as $\polynomialdesign := \left\{ A(p) \;|\; \text{$p$ is a degree-$d$ monic polynomial over $\mathbb{Z}/q\mathbb{Z}$} \right\}$,
	where $\forall p$, $A(p) :=$ $\{$$(i, p(i))$ $\mid$ $0\leq i \leq k - 1$$\}$.
	%
	Then, for every authentication threshold $\mu$, 
	a custody scheme can be induced by $\polynomialdesign$ and $\mu$.
\end{construction}

It is easy to verify that $\polynomialdesign$ consists of $m$ distinct groups,
and the intersection of any two distinct groups in $\polynomialdesign$ is strictly bounded by $d$ 
by the Fundamental Theorem of Algebra,
i.e.: 
\begin{equation}\label{ieq:polynomial-design}
	\forall A_p,A_q\in\polynomialdesign, A_p\ne A_q
	\implies |A_p\cap A_q| < d.
\end{equation}

Hence, the efficiency factor $\eta$ of the custody scheme induced by polynomial design is lower bounded as below. The proof is present in Appendix~\ref{app:proof-thm-polynomial-design}.
\begin{theorem}~\label{thm:polynomial-design}
	Given parameters $k, q, d$, $n = kq$, $\mu$ and corresponding $r$, the efficiency factor $\eta$ of the custody scheme induced by $\polynomialdesign$ and $\mu$ against a $\gamma$-adversary is lower bounded as follows:
	\[\eta \geq \gamma^{1 - d} \cdot \left.\binom{r}{d} \middle/ {\binom{k}{d}}\right. - 1.\]
\end{theorem}

\begin{figure}[!t]
	\centering
	\includegraphics[width=\textwidth]{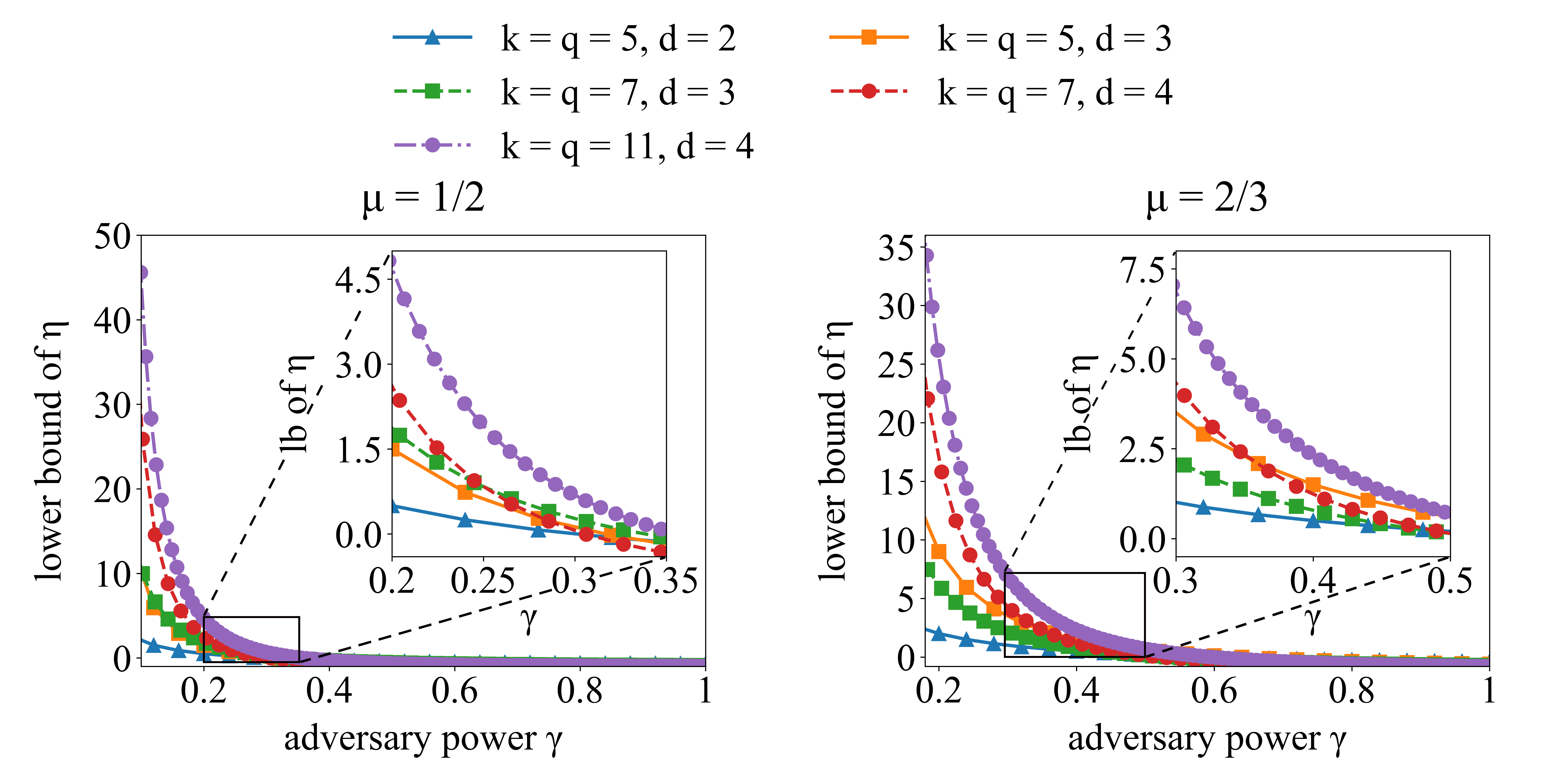}
	\caption{The lower bound of efficiency factor $\eta$ (by Theorem~\ref{thm:polynomial-design}) against adversary power $\gamma$ for $\polynomialdesign$ as in Construction~\ref{constr:polynomial-design}. Recall that $n = kq$ in $\polynomialdesign$.
	}
	\label{fig:polynomial-design}
\end{figure}

Surprisingly, the lower bound of $\eta$ given by Theorem~\ref{thm:polynomial-design} does not rely on the selection of $q$.

From Theorem~\ref{thm:polynomial-design}, we immediately obtain the following proposition:
\begin{proposition}\label{prop:polynomial-design}
	Given parameters $k, q, d$, $n = kq$, $\mu$ and corresponding $r$, the custody scheme induced by $\polynomialdesign$ and $\mu$ is secure against $\gammapolynomialdesign$-adversary for $\gammapolynomialdesign$ defined as:
	\[\gammapolynomialdesign := \left(\binom{r}{d} \middle/ \binom{k}{d}\right)^{\frac{1}{d - 1}}.\]	
\end{proposition}

Fig.~\ref{fig:polynomial-design} depicts the relation between the lower bound of $\eta$ following Theorem~\ref{thm:polynomial-design} against the adversary power $\gamma$,
for $\mu \in \{1/2, 2/3\}$ and $k,q,d$ as shown in the figure. 
Note that $n = kq$.
It is easy to see that the lower bound of $\eta$ increases as $k$ and $d$ become larger with fixed corrupted fraction $\gamma$.
For specific choices we get $\eta \geq 9.45$ against adversary with $\gamma = 3/11$, 
when $\mu = 2/3$ and $\polynomialdesign$ is parameterized by $n = 121$, $k = q = 11$ and $d = 4$, with totally $m = 14,641$ groups. 
Furthermore, we remark that under the estimation of Theorem~\ref{thm:polynomial-design}, the efficiency factor $\eta$ increases rapidly as $\gamma$ decreases since $\eta \sim \gamma^{1 - d}$. For instance,
the lower bound for $\eta$ is improved to no less than $34.29$ when $\gamma$ is reduced from $3/11$ to $2/11$ in the above example.

The polynomial design only implies a group number of $k^d = O(n^{d/2})$, which is far smaller than the group number of $\binom{n}{k}$ given by symmetric design. Our subsequent experiments (see Appendix~\ref{app:experimental-results}) show that 
considering efficiency factor, polynomial design behaves a bit worse than symmetric design. Nevertheless, the result is pleasing enough for a realization in practice. 

\subsection{Block Design}\label{sec:block-design}

One may notice that the previous two constructions give custody schemes with a rather large number of groups. For symmetric design, we have $\binom{n}{k}$ groups; and for polynomial design, we have $k^d = \Theta(n^{d/2})$ groups. Now we come to constructions which lead to a smaller number of groups. Specifically, in this section, we consider block designs. We also study a multi-layer sharding design in Appendix~\ref{app:multi-layer-sharding-design}. 

A block design is a particular combinatorial design consisting of a set of elements and a family of subsets (called blocks) whose arrangements satisfy generalized concepts of balance and symmetry.

\begin{construction}[Block design, from~\cite{Stinson07}, with notation revised]\label{constr:block-design}
	Let $n, k, \lambda$ and $t$ be positive integers such that $n > k \geq t$. 
	$(S, \blockdesign)$ is called a $t$-$(n, k, \lambda)$-design if $S$ is a set with $|S| = n$ and $\blockdesign$ is a family of $k$-subsets of $S$ (called blocks), such that
	every $t$-subset of $S$ is contained in exactly $\lambda$ blocks in $\blockdesign$. One can verify that the number of blocks of a $t$-$(n, k, \lambda)$-design is $m = \lambda\cdot \left. \binom{n}{t} \middle/ \binom{k}{t} \right.$.
\end{construction}

In fact, block design naturally extends symmetric design (Construction~\ref{constr:symmetric-design}),
in the sense that $\symmetricdesign$ is a degenerated block design with $t = k$ and $\lambda = 1$. 
In what follows, a ``block'' in the block design is also called a ``group'' in the group assignment scheme.

The following theorem, which is proven in Appendix~\ref{app:proof-thm-block-design}, shows the effectiveness of block designs:
\begin{theorem}\label{thm:block-design}
    For every $t$-$(n, k, \lambda)$-design $(S, \blockdesign)$, let $\mu \geq (t - 1)/k$ (which implies that $r \geq t$),
    then the efficiency factor $\eta$ of the custody scheme induced by $\blockdesign$ and $\mu$ against a $\gamma$-adversary (i.e., the adversary corrupting $s = \gamma n$ nodes) is lower bounded as follows:
    \[\eta \geq \gamma\cdot \frac{\binom{n}{t}}{\binom{k}{t}}\cdot \frac{\binom{r}{t}}{\binom{s}{t}} - 1.\]
\end{theorem}

The following proposition further shows that the custody scheme induced by block design is secure with proper $\gamma$. The proof of the proposition is in Appendix~\ref{app:proof-prop-block-design}.
\begin{proposition}\label{prop:block-design}
	When $n \geq 3k - 3$, and $\mu \geq 1/2$, $r\geq \max \{t, 3\}$, the custody scheme induced by an $r$-$(n, k, \lambda)$-design with $\mu$ is secure against $\gammablockdesign$-adversary, for $\gammablockdesign$ defined as follows:
	\[\gammablockdesign := \frac{1}{k} \cdot \mu^{\frac{1}{t - 1}} + \frac{t - 1}{n}.\]
\end{proposition}

\begin{figure}[!t]
	\centering
	\includegraphics[width=\textwidth]{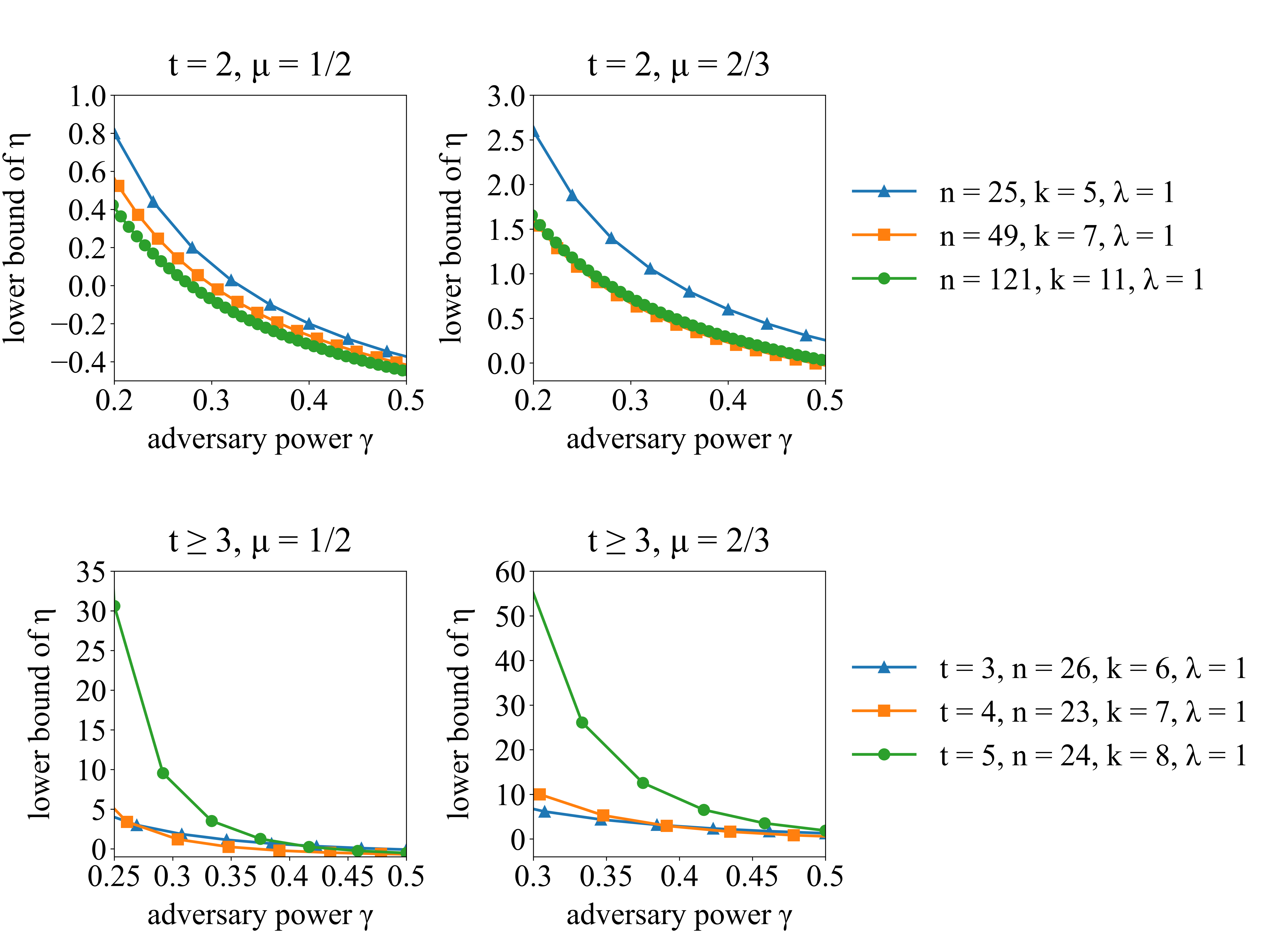}
	\caption{The lower bound of efficiency factor $\eta$ (by Theorem~\ref{thm:block-design}) against adversary power $\gamma$ for $\blockdesign$ as in Construction~\ref{constr:block-design}. All six concrete block designs shown in this figure have explicit constructions~\cite{colbourn2006handbook,Stinson07}.}
	\label{fig:block-design}
\end{figure}

%

When $t = 2$, according to Theorem~\ref{thm:block-design}, we have $\eta \geq \frac{n - 1}{s - 1}\cdot \frac{r(r - 1)}{k(k - 1)} - 1 \approx \frac{\mu^2}{\gamma} - 1$,
which implies that the efficiency factor is at least $\Omega(1)$ when $\gamma \geq 1/2\cdot \mu$. When $k$ and $\lambda$ are constant, the corresponding number of groups is $\lambda\cdot \left.\binom{n}{2} \middle/ \binom{k}{2} \right. = \Theta(n^2)$. With larger $t$, the result given by Theorem~\ref{thm:block-design} is even more inspiring.

Fig.~\ref{fig:block-design} shows the lower bound of $\eta$ obtained by Theorem~\ref{thm:block-design} versus the adversary's power $\gamma$ for different block designs with $\mu \in \{1/2, 2/3\}$. We clearly observe that the lower bound of $\eta$ significantly increases with the value of $t$ under fixed corrupted fraction $\gamma$. Further, although Theorem~\ref{thm:block-design} only provides a lower bound estimation for large $\gamma$, 
we still achieve satisfying numerical results. 
For instance, using the custody scheme induced from the $5$-$(24, 8, 1)$-design (see~\cite{colbourn2006handbook,Stinson07} for the construction) with $m = 759$ custodian groups and $\mu = 1/2$,
the efficiency factor $\eta$ is no less than $30.62$ when $\gamma \le 1/4$.

Meanwhile, our further experimental results demonstrate that block design has a comparable performance with polynomial design (See Appendix~\ref{app:experimental-results}), which indicates that block design finds its application in constructing custodian groups under the scenario of decentralized asset custody in our model.


\section{Compressing Group Assignment Schemes via Random Sampling}\label{sec:random-sampling}

We notice that under symmetric design and polynomial design, a group assignment scheme $\mathcal{A}$ may contain too many custodian groups, which renders the induced custody scheme almost impossible to manage in practice.
To mitigate this problem, 
we propose a randomized sampling technique to 
construct compact custody schemes with a smaller number of custodian groups sampled from $\mathcal{A}$ as representatives.

\begin{construction}[Random sampling]\label{constr:random-sampling}
	Given a group assignment scheme $\mathcal{A}$ consisting of $m$ groups,
	as well as a sampling rate $\beta \in (0,1)$, 
	we uniformly sample a subset of $\beta m$ elements from $\mathcal{A}$ at random as the new assignment scheme $\mathcal{A}'$,
	and then construct a custody scheme based on $\mathcal{A}'$. The sampling process does not affect on the authentication threshold $\mu$.
\end{construction}

In what follows we analyze the efficiency of $\mathcal{A}'$ comparing to $\mathcal{A}$. For a given corrupted fraction $\gamma$, let $H(\gamma)$ be a function of $\gamma$ defined as $H(\gamma) := -\left(\gamma \ln \gamma + \left(1 - \gamma\right)\ln (1 - \gamma)\right)$. Then the efficiency factor of custody scheme induced by $\mathcal{A}'$ is lower bounded as in the following theorem, which is proven in Appendix~\ref{app:proof-thm-random-sampling}:

\begin{theorem}\label{thm:random-sampling}
	Let $\mathcal{A}$ and $\mathcal{A}'$ be defined as above,
	and suppose the corrupted fraction $\gamma$ satisfies $n \gamma(1 - \gamma) \geq 1$.\footnote{This is trivial if $n > 4$ and $\gamma n\ge 2$.} 
	Let $\eta$ and $\eta'$ be the efficiency factor of the custody scheme induced by respectively $\mathcal{A}$ and $\mathcal{A'}$ together with some fixed $\mu$ against a $\gamma$-adversary.
	Then, for arbitrary $c \geq 0$,  with probability at least $1 - \frac{e}{2\pi}\exp(-cnH(\gamma))$,  the following lower bound for $\eta'$ holds: 
	\[\eta' \geq \frac{\gamma(\eta + 1)\cdot \sqrt{\beta m}}{\gamma \cdot\sqrt{\beta m} + (\eta + 1)\cdot \sqrt{(1 + c)n H(\gamma) / 2}} - 1.\]
\end{theorem}

For $c = 1$, Theorem~\ref{thm:random-sampling} transforms into an easy-to-digest version as in 
Corollary~\ref{coro:random-sampling}.
\begin{corollary}\label{coro:random-sampling}
	Let $\eta$ be the efficiency factor of the custody scheme induced by $\mathcal{A}$ and some $\mu$ against a $\gamma$-adversary.
	Let $\mathcal{A'}$ be the group assignment scheme uniformly sampled from $\mathcal{A}$ at random with $m'$ groups. Suppose the custody scheme induced by $\mathcal{A'}$ and $\mu$ has efficiency factor $\eta'$ against the same $\gamma$-adversary.
	Then, with probability at least $1 - \frac{e}{2\pi}\exp(-nH(\gamma))$,
	\begin{itemize}
		\item $\eta' \geq \sqrt{\eta + 1} - 2$, with $m' = (\eta + 1)nH(\gamma)/\gamma^2$;
		\item $\eta' \geq (\eta - 1)/2$, with $m' = (\eta + 1)^2nH(\gamma)/\gamma^2$.
	\end{itemize}
\end{corollary}

To better illustrate the effect of Corollary~\ref{coro:random-sampling}, we consider the symmetric design in Section~\ref{sec:symmetric-design}. (\ref{ieq:symmetric-design}) shows that the efficiency factor of the custody scheme induced by symmetric design reaches $\Theta(n)$ with $k = \Theta(\log n)$, and $\Theta(1)$ with $k = \Theta(1)$. Combining with Corollary~\ref{coro:random-sampling}, we further obtain the following important corollary:

\begin{corollary}\label{coro:key}
	For fixed $\gamma < \mu$, we can uniformly choose $m$ different $k$-subsets of $S$ at random, where $|S| = n$, such that with probability $1 - O(\exp(-nH(\gamma)))$, the efficiency factor $\eta$ of the custody scheme induced by these subsets and $\mu$ against a $\gamma$-adversary satisfies:
	\begin{itemize}
		\item $\eta = \Omega(1)$, with $k = \Theta(1)$ and $m = \Theta(n)$;
		\item $\eta = \Omega(\sqrt{n})$, with $k = \Theta(\log n)$ and $m = \Theta(n^2)$;
		\item $\eta = \Omega(n)$, with $k = \Theta(\log n)$ and $m = \Theta(n^3)$.
	\end{itemize}
\end{corollary}

\section{Summary and Discussion}\label{sec:discussion}

%

In this work we propose a framework of decentralized asset custody schemes based on overlapping group assignments.
The custody scheme reaches high efficiency, with security guaranteed against any rational adversary that corrupts a bounded fraction of custodians.

Explicit constructions of compact assignments with much less custodian groups,
efficient approximation algorithms for estimating the actual efficiency factor of a given custody scheme in our framework, 
and more rigorous analysis of liveness guarantee as well as the trade-off between liveness and security are of independent interest, which we left for future work.

\bibliographystyle{splncs04}
\bibliography{References}

\newpage

\begin{subappendices}
\renewcommand{\thesection}{\Alph{section}}

\section{Multi-layer Sharding Design}\label{app:multi-layer-sharding-design}

In this section, we consider a multi-layer sharding design, which is an extension of the simple idea to partition all nodes into non-overlapping groups with an identical size.
\begin{construction}[Multi-layer sharding design]\label{constr:multi-layer-sharding-design}
	For given $k$, let the number of custodians $|S| = n$ be a multiple of $k$. We use \emph{sharding layer} to term a partition of $S$ into $n/k$ groups, each with $k$ custodians. Let $l$ be any positive integer. With $n, k, l$, a multi-layer sharding design $\multilayershardingdesign$ consists of $l$ sharding layers of $S$. Notice that such a group assignment scheme may include some groups more than once. In particular, we use $\randommultilayershardingdesign$ to specify the realization of multi-layer sharding design such that each sharding layer is independently and uniformly drawn from all possibilities.
\end{construction}

We are specifically interested in analyzing the custody scheme induced by $\randommultilayershardingdesign$. With $n, k, l$ given, for some adversary power $\gamma$ and corresponding number of currupted nodes, resembling Section~\ref{sec:random-sampling}, we let
\[H(\gamma) := -\left(\gamma \ln \gamma + \left(1 - \gamma\right)\ln (1 - \gamma)\right).\]
also, we let
\[\kappa(\gamma) := \Pr[\mathcal{H}(n, s, k) \geq r].\]
Further, with any $\delta > 0$, we define
\[\tau(\gamma, \delta) := \frac{2\delta^2\cdot n^2 / k^2 \cdot l \cdot {\kappa(\gamma)}^2}{n\cdot \left(s / r - s / k + 1\right)^2} = \frac{l}{n}\cdot \frac{2\delta^2\cdot  {\kappa(\gamma)}^2}{\gamma^2\cdot(1 / r - 1/ k + 1/ s)^2\cdot k^2}.\]

We have the following theorem:
\begin{theorem}\label{thm:random-multi-layer-sharding-design}
	Given parameters $n, k, l$, $\mu$ and corresponding $r$, the efficiency factor $\eta$ of the custody scheme induced by $\randommultilayershardingdesign$ and $\mu$ against a $\gamma$-adversary satisfies
	\[\Pr\left[\eta \geq \frac{\gamma}{(1 + \delta)\cdot \kappa(\gamma)} - 1\right] \geq 1 - \frac{e}{2\pi}\cdot \frac{1}{\sqrt{n\gamma(1 - \gamma)}}\cdot \exp(-n\cdot (\tau(\gamma, \delta) - H(\gamma))).\]
\end{theorem}

\begin{proof}
	First, consider a fixed set $P$ of $s = \gamma n$ nodes that are corrupted by the adversary. For $1\leq i \leq l$, let $X_i$ be a random variable denoting the number of corrupted groups in the $i$-th sharding layer, and $X = X_1 + \cdots + X_l$ be the total number of corrupted groups. 
	Clearly, the following inequalities hold:
	\[\frac{s}{k} - 1 \leq X_i \leq \frac{s}{r}, \quad\forall 1\leq i \leq l.\]
	
	Meanwhile, according to the definition of $\randommultilayershardingdesign$, we know that $X_1, \cdots, X_l$ are i.i.d with
	\[\E[X_i] = \frac{n}{k}\cdot \kappa(\gamma), \quad\forall 1\leq i \leq l.\]
	Therefore,
	\[\E[X] = \E\left[\sum_{i = 1}^{l}X_i\right] = \sum_{i = 1}^{l}\E\left[X_i\right] = l\cdot \frac{n}{k}\cdot \kappa(\gamma).\]
	
	By Hoeffding's inequality~\cite{Hoeffding63}, we know that for any $\delta > 0$, 
	\begin{align*}
		\Pr[X \geq (1 + \delta)\E[X]\mid\mathcal{Z}] &\leq \exp \left(-\frac{2\delta^2\cdot n^2 / k^2 \cdot l \cdot {\kappa(\gamma)}^2}{\left(s / r - s / k + 1\right)^2}\right) \\
		&= \exp \left(-n\cdot \tau(\gamma, \delta)\right).
	\end{align*}
	
	With regard to all $\binom{n}{s}$ possibilities of $P$ the corrupted set, by a union bound and Stirling's formula, we obtain that
	\begin{align*}
		&\Pr[\maxcorruptedgroupnumberrmls \geq (1 + \delta)\E[X]] \\ \leq\ &\binom{n}{s}\cdot \exp \left(-n\cdot \tau(\gamma, \delta)\right) \\
		\leq\ &\frac{e}{2\pi}\cdot \frac{1}{\sqrt{n\gamma(1 - \gamma)}}\cdot \exp(-n\cdot (\tau(\gamma, \delta) - H(\gamma))).		
	\end{align*}
	
	Finally, when $\maxcorruptedgroupnumberrmls \geq (1 + \delta)\E[X]$ holds, we further derive that
	\[\eta = \frac{\gamma\cdot m}{\maxcorruptedgroupnumberrmls} - 1 \geq \frac{\gamma}{(1 + \delta)\cdot \kappa(\gamma)} - 1,\]
	which finishes the proof.
	\qed
\end{proof}

It is worth noting that in Theorem~\ref{thm:random-multi-layer-sharding-design}, in order that the failure probability is negligible in $n$, it is required that $\tau(\gamma, \delta) = \Omega(1)$. As per definition, $\tau(\gamma, \delta) = \Theta(l/n)$ when $k$ is a constant. Therefore, we demand that $l = \Omega(n)$, or the total number of groups $m = \Omega(n^2$).

\begin{figure}[!t]
	\centering
	\includegraphics[width=\textwidth]{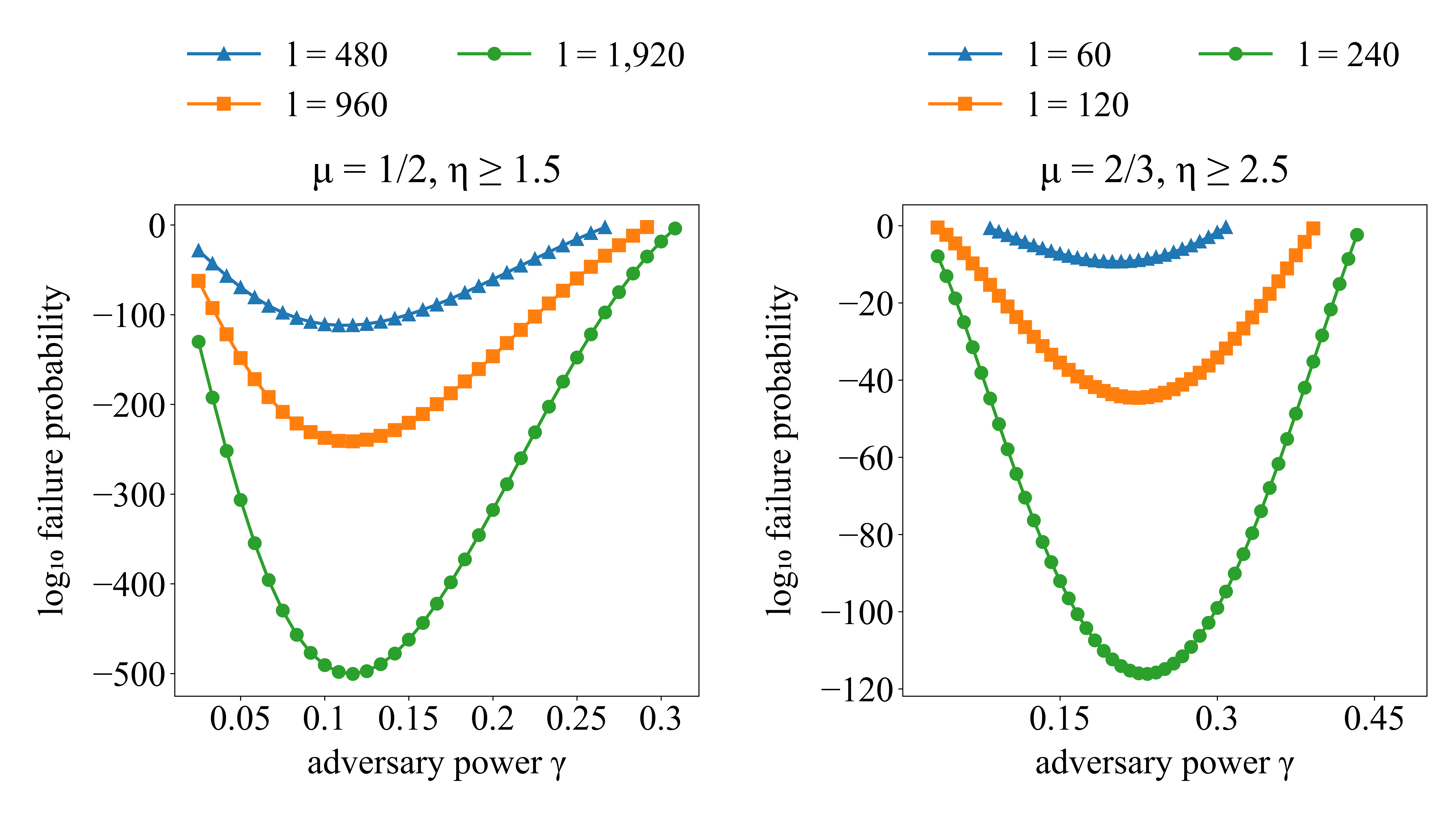}
	\caption{The logarithm of failure probability to the base $10$ (by Theorem~\ref{thm:random-multi-layer-sharding-design}) of efficiency factor $\eta$ no less than given value against adversary power $\gamma$ for $\randommultilayershardingdesign$ with $n = 120$ and $k = 5$ as in Construction~\ref{constr:multi-layer-sharding-design}. }
	\label{fig:rmls-design}
\end{figure}

Fig.~\ref{fig:rmls-design} shows the logarithm of failure probability to the base $10$ obtained by Theorem~\ref{thm:block-design}, that $\eta$ is no less than $1.5$ or $2.5$ when $\mu = 1/2$ or $\mu = 1/3$ for random multi-layer sharding designs with $n = 120$ and $k = 5$. We see that with the accumulation of layers, the failure probability shows an exponential decrease. In general, random multi-layer sharding designs turn out to be acceptable. Concretely, when $n = 120$, $k = 5$ and $l = 480$, the efficiency factor of the custody scheme induced by $\randommultilayershardingdesign$ and $\mu = 1/2$ is no less than $1.5$ with probability $1 - 10^{-15}$ for adversary power $\gamma = 1/4$. When $l = 60$ and $\mu = 2/3$, such failure is enhanced to $1 - 10^{-19}$ for a higher threshold $\eta \geq 2.5$. Nevertheless, our empirical results reveal that random multi-layer sharding design behaves worse than polynomial design and block design, both on efficiency factor and the number of groups. See Appendix~\ref{app:experimental-results} for details.

\section{Hardness Results}\label{app:hardness}
In this section, we consider the hardness issue of finding the best corrupting strategy, given a group assignment scheme and the number of corrupted nodes. In general, we show that such problem is NP-hard, in the following theorem.

\begin{theorem}\label{thm:hardness}
	Suppose $1\leq r \leq k$ are given constants, where $r = \lceil \mu k + \epsilon \rceil$. Let $|S| = n$, and $\mathcal{A}$ be a group assignment scheme such that $m = |\mathcal{A}|$ is within a polynomial size of $n$, and all groups in $\mathcal{A}$ have size $k$. Further, suppose an adversary controlling $r < s = \gamma n < n$ nodes. Then, when $1 \leq r < k$, it is NP-hard for the adversary to find the optimal attacking strategy and compute the value $\maxcorruptedgroupnumber$.
\end{theorem}
\begin{proof}
	We first deal with the case when $r = 1$. In fact, in this case, when $k = 2$, the origin problem is equivalent to MAX $s$-VC in common graphs, which is known to be NP-hard~\cite{CornuejolsNW80}: 
	\begin{problem}[MAX $s$-VC]\label{prob:max-s-vc}
		Given a common graph $G = (V, E)$, $|V| = n$, and $s = n$, find a subset $P$ of $s$ vertices that maximizes the total number of edges covered by $P$. We say an edge is covered by $P$ iff at least one of its endpoints lies in $P$.
	\end{problem} 
	
	When $r = 1$ and $k > 2$, the problem is equivalent to the MAX $s$-VC on $k$-uniform hypergraph. Here, a $k$-uniform hypergraph is a hypergraph in which each edge contains exactly $k$ vertices.
	\begin{problem}[MAX $s$-VC on $k$-uniform hypergraph]\label{prob:max-s-vc-on-k-uniform-hypergraph}
		Given a $k$-uniform hypergraph $G = (V, E)$, $|V| = n$, and $s = n$, find a subset $P$ of $s$ vertices that maximizes the total number of hyperedges covered by $P$. We say a hyperedge is covered by $P$ iff at least one of its endpoints lies in $P$.
	\end{problem}
	
	We reduce MAX $s$-VC in common graphs to this problem. Specifically, consider a realization of the problem in the common graph $G = (V, E)$ with $m$ edges. We transfer $G$ to a $k$-uniform hypergraph $G' = (V', E')$ by adding $(k - 2)m$ vertices to $V$. Specifically, denote these vertices as $v_1^1, v_1^2, \cdots, v_1^m$, $\cdots$, $v_{k - 2}^1, v_{k - 2}^2, \cdots, v_{k - 2}^m$. For each edge $e_i$ in $G$, $1\leq i \leq m$, we add $v_1^i, v_2^i, \cdots, v_{k - 2}^i$ to $e_i$ and obtain an edge $e_i'$ in $E'$ containing $k$ vertices. 
	
	Now consider the MAX $s$-VC solution in $k$-uniform hypergraph $G'$. We show that there is an optimum that contains no vertex in $V'\setminus V$. In fact, we suppose that there is an optimal solution containing some vertex $v$ in $V'\setminus V$. Note that $v$ is incident with only one hyperedge $e$. On one hand, if not both two vertices in $e$ that belong to $V$ are selected, then we can replace $v$ with any unselected vertex in these two vertices, still remaining an optimal solution. On the other hand, if both vertices in that edge that belong to $V$ are selected, we can alternatively pick any unselected vertex in $V$, still obtaining an optimal solution. We can always achieve this, as $s < n$. According to such a method, we can successively substitute all selected vertices in $V'\setminus V$ with vertives in $V$, eventually achieving an optimal solution with selected vertices all in $V$. Therefore, the optimal solution in the $k$-uniform hypergraph immediately leads to the optimal solution in the original problem instance in the common graph, by ignoring all vertices in $V'\setminus V$. Note that the reduction runs in polynomial time since $m$ is within a polynomial size of $n$ and $k$ is a constant. Hence, MAX $s$-VC on $k$-uniform hypergraph is NP-hard as well.
	
	We should mention that the MAX $s$-VC problem in $k$-uniform hypergraphs ($k \geq 2$) is indeed a particular case of MAX $s$-Cover, with each element existing in precisely $k$ sets. The reduction is to deem each vertex as a set, including all edges incident to the vertex. There is a simple $(1 - e^{-1})$-approximation polynomial-time greedy algorithm for MAX $s$-Cover~\cite{Hochbaum96,Cornuejols77}. Furthermore, it is known that in general, for any $\epsilon > 0$, there is no deterministic $(1 - e^{-1} + \epsilon)$-approximation for this problem in polynomial time unless P = NP~\cite{Feige98}.
	
	When $k > r > 1$, the original problem is equivalent to the following problem:
	\begin{problem}[MAX $s$-Vertex $r$-Cover on $k$-uniform hypergraph]\label{prob:max-s-v-r-c-on-k-uniform-hypergraph}
		Given a $k$-uniform hypergraph $G = (V, E)$, $|V| = n$, and $s = n$, find a subset $P$ of $s$ vertices that maximizes the total number of hyperedges $r$-covered by $P$. We say a hyperedge is $r$-covered by $P$ iff at least $r$ of its endpoints lies in $P$.
	\end{problem}
	
	We reduce MAX $(s - r + 1)$-VC in $(k - r + 1)$-uniform hypergraph to this problem. Again, consider an instance $G = (V, E)$ of MAX $(s - r + 1)$-VC in $(k - r + 1)$-uniform hypergraph. We add $(r - 1)$ vertices to $V$, as well as including them in each hyperedge in $E$ to achieve $G' = (V', E')$, which is an instance of Problem~\ref{prob:max-s-v-r-c-on-k-uniform-hypergraph}. Consider the optimal solution in this instance. We claim that there always exists an optimal solution that includes the appended $(r - 1)$ vertices. To show this, consider any optimal solution with some vertex $v$ in $V'\setminus V$ unselected. Then, replacing any selected vertex in $V$ with $v$ also achieves optimal, as $v$ is contained in all hyperedges. Therefore, under such optimal solution, an edge in $G'$ is covered if and only if the corresponding edge in $G$ is covered in the origin MAX $(s - r + 1)$-VC instance in $(k - r + 1)$-uniform hypergraph. Consequently, such an optimal solution in our created instance immediately leads to an optimal solution in the origin problem, by ignoring all vertices in $V'\setminus V$. Furthermore, such reduction runs in polynomial time of $n$, which leads to the NP-hardness of Problem~\ref{prob:max-s-v-r-c-on-k-uniform-hypergraph}.
	\qed
\end{proof}

Theorem~\ref{thm:hardness} shows that generally, it is hard for the computationally bounded adversary to figure out the best attacking strategy. Nevertheless, there is a deterministic algorithm for the adversary to figure out a strategy which corrupts at least an average amount of groups. Resembling Appendix~\ref{app:multi-layer-sharding-design}, let $\kappa(\gamma) := \Pr[\mathcal{H}(n, s, k) \geq r]$. Apparently, $\kappa(\gamma) m$ is the expected number of groups that the adversary can corrupt when the $\gamma n$ corrupted nodes are chosen uniformly at random. We have the following theorem:
\begin{theorem}\label{thm:better-than-average}
	Let $\kappa(\gamma)$ be defined as above. Suppose computing each binomial coefficient with size $n$ takes time no more than $T(n)$. Then there is a deterministic algorithm that gives an attacking strategy which corrupts at least $\kappa(\gamma) m$ groups, running in time $O\left(kmnT(n)\right)$.
\end{theorem}

\begin{proof}
	Algorithm~\ref{alg:better-than-average} shows how to find the desired strategy. Specifically, in step $1 \leq i \leq n$ (Line~\ref{algline:bta-iter}), given a temporary corrupted list $P$ and honest list $Q$ ($P\cup Q = \{1, 2, \cdots, i - 1\}$), the adversary decides whether or not to corrupt node $i$. To figure this out, the adversary needs to compute the expected number of corrupted groups $x_i$ conditioning on nodes in $P \cup \{i\}$ are already corrupted and nodes in $Q$ are honest, and other nodes are corrupted uniformly at random under the constraint that $s$ nodes are corrupted in total (Line~\ref{algline:bta-compute}). In detail, the adversary should compute the conditional probability on each group is corrupted and take a sum over all groups to derive $x_i$. If $x_i \geq \kappa(\gamma) m$, node $i$ will be included in $P$ (Line~\ref{algline:bta-include}), otherwise it will be given up by the adversary (Line~\ref{algline:bta-discard}). The algorithm ends whenever $s$ nodes are already chosen (Line~\ref{algline:bta-end1}) or $n - s$ nodes are already given up (Line~\ref{algline:bta-end2}). Algorithm~\ref{alg:better-than-average} runs in time $O\left(kmnT(n)\right)$ as there are at most $n$ steps, and in each step, one needs to compute the conditional probability for each of $m$ groups, and computing each conditional probability gives the time complexity of $O\left(kT(n)\right)$. 
	\begin{algorithm}[!ht]
		\caption{Corrupting at least an average amount of groups.}
		\label{alg:better-than-average}
		\renewcommand{\algorithmicrequire}{\textbf{Input:}}
		\renewcommand{\algorithmicensure}{\textbf{Output:}}
		\begin{algorithmic}[1]
			\REQUIRE The node set $S = \{1, 2, \cdots, n\}$, the group assignment scheme $\mathcal{A}$ with parameters $m$ and $k$, least number of nodes to control a group $r$, number of corrupted nodes $s$.
			\ENSURE $P \subseteq S$ with $|P| = s$, such that when $P$ is corrupted, the adversary can at least control $\kappa(\gamma) m$ groups, with $\kappa(\gamma)$ defined as $\kappa(\gamma) := \Pr[\mathcal{H}(n, s, k) \geq r]$.
			
			\STATE $P \gets \emptyset, Q \gets \emptyset$
			\FOR{$i \gets 1$ to $n$} \label{algline:bta-iter}
			\STATE Compute the expected number of corrupted groups $x_i$ conditioning on nodes in $P \cup \{i\}$ are corrupted and nodes in $Q$ are honest, and other nodes are corrupted uniformly at random under the constraint that $s$ nodes are malicious in total \label{algline:bta-compute}
			\IF{$x_i \geq \kappa(\gamma) m$} \label{algline:bta-judge}
			\STATE $P \gets P \cup \{i\}$ \label{algline:bta-include}
			\ELSE 
			\STATE $Q \gets Q \cup \{i\}$ \label{algline:bta-discard}
			\ENDIF
			
			\IF{$|P| \geq s$} \label{algline:bta-end1}
			\RETURN $P$
			\ENDIF
			\IF{$|Q| \geq n - s$} \label{algline:bta-end2}
			\RETURN $S\setminus N$
			\ENDIF
			\ENDFOR
		\end{algorithmic}
	\end{algorithm}
	
	Clearly, Algorithm~\ref{alg:better-than-average} is sure to end up with a size-$s$ subset $P$ of $S$. To show that corrupting $P$ leads to at least $\kappa(\gamma) m$ groups controlled by the adversary, we need the following lemma.
	\begin{lemma}\label{lem:better-than-average}
		Suppose Algorithm~\ref{alg:better-than-average} does not end after step $i$. Let $x_i (i \geq 0)$ be the expected amount of corrupted nodes conditioning on $P$ is corrupted, $Q$ is honest after step $i$ and other nodes are corrupted uniformly at random under the constraint that $s$ nodes are corrupted in total, then $x_{i} \geq x_{i - 1}$. Specifically, we have $x_0 = \kappa(\gamma) m$.
	\end{lemma}
	\begin{proof}[Lemma~\ref{lem:better-than-average}]
		To show this lemma, let $Y$ be a random variable denoting the number of corrupted groups. Furthermore, let $a_i < s$ and $b_i < n - s$ be respectively the size of $P$ and $Q$ after step $i$ (denote by the state of $P$ and $Q$ of that time by $P_i$ and $Q_i$, respectively), $a_i + b_i = i$. Then we have the following equality:
		\begin{align*}
			x_i &= \E[Y\mid P_i, Q_i] \\
			&= \frac{s - a_i}{n - i}\cdot \E[Y\mid P_i \cup \{i + 1\}, Q_i] + \frac{n - s - b_i}{n - i}\cdot \E[Y\mid P_i, Q_i \cup \{i + 1\}].
		\end{align*}
		Here, $\E[Y\mid P_i, Q_i]$ is for the expectation of $Y$ conditioning on $P_i$ malicious and $Q_i$ honest. Hence, at least one of the $\E[Y\mid P_i \cup \{i + 1\}, Q_i]$ and $\E[Y\mid P_i, Q_i \cup \{i + 1\}]$ is no less than $x_i$. According to Line~\ref{algline:bta-judge}, $x_{i + 1} \geq x_i$.
		\qed
	\end{proof}
	
	With Lemma~\ref{lem:better-than-average}, note that $x_0 = \kappa(\gamma) m$, the theorem is proved.
	\qed
\end{proof}

\section{Proof of Theorems and Propositions}\label{app:proofs}

\subsection{Proof of Proposition~\protect{\ref{prop:symmetric-design-1}}}\label{app:proof-prop-symmetric-design-1}
\begin{proof}
	Recall that the equivalent condition for the custody scheme induced by the symmetric design to be secure is that
	\[\maxcorruptedgroupnumbersym \leq \gamma\cdot m.\]
	
	Specifically, for symmetric design, we have $m = \binom{n}{k}$, and
	\[\maxcorruptedgroupnumbersym = \binom{n}{k} \cdot \sum_{t = r}^{k} \frac{\binom{\gamma n}{t}\binom{n - \gamma n}{k - t}}{\binom{n}{k}}.\]
	
	Therefore, the scheme is secure iff
	\[\gamma \geq \sum_{t = r}^{k} \frac{\binom{\gamma n}{t}\binom{n - \gamma n}{k - t}}{\binom{n}{k}}.\]
	
	Let $\bar{\mu} := r/k > \gamma$. The tail bound of hypergeometric distribution~\cite{chvatal79} shows that
	\[\sum_{t = r}^{k} \frac{\binom{\gamma n}{t}\binom{n - \gamma n}{k - t}}{\binom{n}{k}} \leq \left(\left(\frac{\gamma}{\bar{\mu}}\right)^{\bar{\mu}}\left(\frac{1 - \gamma}{1 - \bar{\mu}}\right)^{1 - \bar{\mu}}\right)^k.\]
	
	As a result, it is sufficient for the custody scheme to be secure if 
	\[\gamma \geq \left(\left(\frac{\gamma}{\bar{\mu}}\right)^{\bar{\mu}}\left(\frac{1 - \gamma}{1 - \bar{\mu}}\right)^{1 - \bar{\mu}}\right)^k,\]
	or equivalently, 
	\[(r - 1)\ln \gamma + (k - r)\ln(1 - \gamma) \leq r\ln r + (k - r)\ln (k - r) - k\ln k.\]
	
	Now let $\gamma = \frac{r - 1 - x}{k - 1}$, where $0 < x < \min \{r - 1, k - r\}$ is a real number to be determined. The above inequality becomes
	\begin{align*}
		&(k - r)\ln\left(1 + \frac{x}{k - r}\right) + (r - 1)\ln\left(1 - \frac{x}{r - 1}\right) \\
		\leq\ &r\ln r + (k - 1)\ln (k - 1) - (r - 1)\ln(r - 1) - k \ln k.
	\end{align*}
	
	Notice that for any $0 < z < 1$, we have $\ln(1 + z) \leq z$ and $\ln(1 - z) \leq -z - z^2 / 2$. Plugging the result into the above inequality, we derive that it is sufficient if 
	\[-\frac{x^2}{2(r - 1)} \leq r\ln r + (k - 1)\ln (k - 1) - (r - 1)\ln(r - 1) - k \ln k,\]
	or
	\[\frac{x^2}{2(r - 1)} \geq \ln(k - 1) - \ln(r - 1) + r\ln\left(1 - \frac{1}{r}\right) - k\ln\left(1 - \frac{1}{k}\right).\]
	
	We have $r\ln\left(1 - \frac{1}{r}\right) < k\ln\left(1 - \frac{1}{k}\right)$ as $r < k$. Therefore, the custody scheme induced by symmetric design is secure if 
	\[x \geq \sqrt{2(r - 1)\ln \frac{k - 1}{r - 1}},\]
	that is, 
	\[\gamma \leq \frac{r - 1 - \sqrt{2(r - 1)\ln \frac{k - 1}{r - 1}}}{k - 1}.\]
	\qed
\end{proof}

\subsection{Proof of Proposition~\protect{\ref{prop:symmetric-design-2}}}\label{app:proof-prop-symmetric-design-2}
\begin{proof}
	To prove the proposition, the following key lemma is required, which shows that for custody scheme induced by symmetric design, the reliability of the scheme naturally translates into security when $\gamma \leq \min\left\{\frac{r - 1}{k - 1} + \frac{1}{n}, 1 - \frac{k}{n}\right\}$.
	
	\begin{lemma}\label{lem:symmetric-design}
		Given $n$ and $k$, if the custody scheme induced by $\symmetricdesign$ and any $\mu$ is $\gamma$-reliable
		and $\gamma \leq \min\left\{\frac{r - 1}{k - 1} + \frac{1}{n}, 1 - \frac{k}{n}\right\}$,
		then it is secure against $\gamma$-adversary.
	\end{lemma}
	
	\begin{proof}[Lemma~\ref{lem:symmetric-design}]
		Let 
		\[\eta_{s} = \frac{s}{n}\frac{\binom{n}{k}}{\sum_{t = r}^{k}\binom{s}{t}\binom{n - s}{k - t}} - 1\]
		be the efficiency factor of the custody scheme with $s$ corrupted nodes. We will compare $\eta_{s}$ with $\eta_{s - 1}$. Specifically, we compare every corresponding pair of terms in the sum of $1 / (\eta_{s} + 1)$ and $1 / (\eta_{s - 1} + 1)$. For any $r \leq t \leq k$, we have
		\begin{align*}
			&\left. s\cdot \frac{\binom{n}{k}}{\binom{s}{t}\binom{n - s}{k - t}} \middle/ (s - 1)\cdot \frac{\binom{n}{k}}{\binom{s - 1}{t}\binom{n - s + 1}{k - t}} \right. \\
			=\ &\left. \frac{(n - s - k + t)!(s - t)!}{(n - s)!(s - 1)!} \middle/ \frac{(n - s - k + t + 1)!(s - t - 1)!}{(n - s + 1)!(s - 2)!} \right. \\
			=\ &\frac{(s - t)(n - s + 1)}{(n - s - k + t + 1)(s - 1)},
		\end{align*}
		and
		\begin{align*}
			&\frac{(s - t)(n - s + 1)}{(n - s - k + t + 1)(s - 1)} \leq 1 \\
			\Longleftrightarrow\ &(s - t)(n - s + 1) \leq (n - s - k + t + 1)(s - 1) \\
			\Longleftrightarrow\ &s(k - 1) \leq (t - 1)n + k - 1 \\
			\Longleftarrow\ &s \leq \frac{r - 1}{k - 1}n + 1.
		\end{align*}
		Here, the second inequality is due to $s \leq n - k$, while the fourth inequality is due to $t \geq r$. 
		
		Therefore, 
		\begin{align*}
			\frac{(s - t)(n - s + 1)}{(n - s - k + t + 1)(s - 1)} \leq 1 
			&\Longleftrightarrow s\cdot \frac{\binom{n}{k}}{\binom{s}{t}\binom{n - s}{k - t}} \leq (s - 1)\cdot \frac{\binom{n}{k}}{\binom{s - 1}{t}\binom{n - s + 1}{k - t}} \\
			&\Longleftrightarrow \frac{n}{s}\cdot \frac{\binom{s}{t}\binom{n - s}{k - t}}{\binom{n}{k}} \geq \frac{n}{s - 1}\cdot \frac{\binom{s - 1}{t}\binom{n - s + 1}{k - t}}{\binom{n}{k}} \\
			&\Longleftrightarrow \frac{n}{s}\cdot \sum_{t = r}^{k}\frac{\binom{s}{t}\binom{n - s}{k - t}}{\binom{n}{k}} \geq \frac{n}{s - 1}\cdot \sum_{t = r}^{k}\frac{\binom{s - 1}{t}\binom{n - s + 1}{k - t}}{\binom{n}{k}} \\
			&\Longleftrightarrow \frac{1}{\eta_{s} + 1} \geq \frac{1}{\eta_{s - 1} + 1}  \\
			&\Longleftrightarrow \eta_s \leq \eta_{s - 1}.
		\end{align*}
		always holds when $s \leq \frac{r - 1}{k - 1}n + 1$, which proves the lemma as $s = \gamma n$.
		\qed
	\end{proof}
	
	Now for the proposition to prove, notice that under the given conditions, we have
	\begin{align*}
		\maxcorruptedgroupnumbersym &= \binom{n}{k} \cdot \sum_{t = r}^{k} \frac{\binom{s}{t}\binom{n - s}{k - t}}{\binom{n}{k}} \\
		&= \binom{n}{k} \sum_{t = (k + 1) / 2}^{k} \frac{\binom{n / 2}{t}\binom{n / 2}{k - t}}{\binom{n}{k}} \\
		&= \frac{1}{2}\binom{n}{k} \sum_{t = 0}^{k} \frac{\binom{n / 2}{t}\binom{n / 2}{k - t}}{\binom{n}{k}} \\
		&= \frac{1}{2}\binom{n}{k}.
	\end{align*}
	
	As $m = \binom{n}{k}$, the custody scheme is reliable with $\gamma = 1/2$. Further, since $1/2 \leq \min\left\{\frac{r - 1}{k - 1} + \frac{1}{n}, 1 - \frac{k}{n}\right\}$, by Lemma~\ref{lem:symmetric-design}, the proposition is achieved.
	\qed
\end{proof}

\subsection{Proof of Theorem~\protect{\ref{thm:polynomial-design}}}\label{app:proof-thm-polynomial-design}
\begin{proof}
	We say a subset of $T = \{(a, b) \;|\; 0\leq a\leq k - 1, 0\leq b \leq q - 1\}$ is \emph{first-entry-unrepeated} if all nodes in the subset are with different first entry. Now that the adversary corrupts $s$ nodes in total. Let $s_i$ be the number of corrupted nodes with first entry $i$. We have $s = s_0 + s_1 + \cdots s_{k - 1}$. Therefore, the number of size-$d$ first-entry-unrepeated subsets that is totally corrupted is 
	\[\sum_{0 \leq i_1 < i_2 \cdots < i_d \leq k - 1}s_{i_1}s_{i_2}\cdots s_{i_d}.\]
	
	To give an upper bound on the above formula, we extend the problem to the case where $s_0, s_1, \cdots, s_{k - 1}$ are multiples of $1/k$. (The original problem is when $s_0, s_1, \cdots, s_{k - 1}$ are all integers.) For any two indices $u, v$, suppose $s_v - s_u \geq 2/k$, note that 
	\begin{align*}
		&\sum_{0 \leq i_1 < i_2 \cdots < i_d \leq k - 1}s_{i_1}s_{i_2}\cdots s_{i_d} \\
		=\ &s_us_v\sum_{\substack{0 \leq i_1 < i_2 \cdots < i_{d - 2} \leq k - 1
			\\
			\{i_1, i_2, \cdots, i_{d - 2}\} \cap \{u, v\} = \emptyset}}s_{i_1}s_{i_2}\cdots s_{i_{d - 2}} \\
		+\ &(s_u + s_v)\sum_{\substack{0 \leq i_1 < i_2 \cdots < i_{d - 1} \leq k - 1
			\\
			\{i_1, i_2, \cdots, i_{d - 1}\} \cap \{u, v\} = \emptyset}}s_{i_1}s_{i_2}\cdots s_{i_{d - 1}} \\
		+\ &\sum_{\substack{0 \leq i_1 < i_2 \cdots < i_d \leq k - 1
			\\
			\{i_1, i_2, \cdots, i_d\} \cap \{u, v\} = \emptyset}}s_{i_1}s_{i_2}\cdots s_{i_d}.
	\end{align*}
	
	Hence, the sum is strictly increase by substituting $s_u$ with $s_u' = s_u + 1/k$ and $s_v$ with $s_v' = s_v - 1/k$. Therefore, the sum reaches a maximum with $s_0 = s_1 = \cdots = s_{k - 1} = s/k$ in the generalized case, which implies that with $s_0, s_1, \cdots, s_{k - 1}$ all integers, we have
	\[\sum_{0 \leq i_1 < i_2 \cdots < i_d \leq k - 1}s_{i_1}s_{i_2}\cdots s_{i_d} \leq \binom{k}{d}\left(\frac{s}{k}\right)^d,\]
	or that the number of size-$d$ first-entry-unrepeated subsets that is totally controlled by the adversary is upper bounded by $\binom{k}{d} (s / k)^d$. Let $P$ denote the set of corrupted nodes. By (\ref{ieq:polynomial-design}), every size-$d$ first-entry-unrepeated subset of $P$ appears in at most one corrupted group. On the other hand, every corrupted group contains at least $r$ corrupted nodes all with different first entry, and hence $\ge \binom{r}{d}$ size-$d$ first-entry-unrepeated subsets. Consequently, the number of corrupted groups $\maxcorruptedgroupnumberpoly$ is upper bounded by $\frac{\binom{k}{d}s^d}{\binom{r}{d}k^d}$, and the efficiency factor $\eta$ is lower bounded by
	\begin{align*}
		\eta = \frac{\gamma\cdot m}{\maxcorruptedgroupnumberpoly} - 1 &\geq  \gamma\cdot \left.\left(\frac{k}{s}\right)^d\cdot q^d\cdot \binom{r}{d}\middle/\binom{k}{d}\right. - 1 \\
		&= \gamma^{1 - d}\cdot \left.\binom{r}{d} \middle/ {\binom{k}{d}}\right. - 1.
	\end{align*}
	\qed
\end{proof}

\subsection{Proof of Theorem~\protect{\ref{thm:block-design}}}\label{app:proof-thm-block-design}
\begin{proof}
	We say a subset of $S$ is completely corrupted, if all nodes in the subset are corrupted. Notice that for any group, it is corrupted only when at least $\binom{r}{t}$ size-$t$ subsets of the group are completely corrupted. At the same time, when exactly $s$ nodes are corrupted, the adversary can completely corrupt $\binom{s}{t}$ size-$t$ subsets of $S$. As any size-$t$ subset of $S$ appears in exactly $\lambda$ groups. Therefore, we have
	\[\maxcorruptedgroupnumberblck \leq \lambda\cdot \left.\binom{s}{t} \middle/ \binom{r}{t}\right..\]
	
	Further notice that $m = \lambda\cdot \left. \binom{n}{t} \middle/ \binom{k}{t} \right.$, we have
	\[\eta = \frac{\gamma\cdot m}{\maxcorruptedgroupnumberblck} - 1\geq \gamma\cdot \frac{\binom{n}{t}}{\binom{k}{t}}\cdot \frac{\binom{r}{t}}{\binom{s}{t}} - 1. \]
	\qed
\end{proof}

\subsection{Proof of Proposition~\protect{\ref{prop:block-design}}}\label{app:proof-prop-block-design}
\begin{proof}
	By Theorem~\ref{thm:block-design}, we have the following lower bound on the efficiency factor of the custody scheme:
	\[\eta \geq \gamma\cdot \frac{\binom{n}{t}}{\binom{k}{t}}\cdot \frac{\binom{r}{t}}{\binom{s}{t}} - 1.\]
	
	For the custody scheme to be $\gamma$-reliable, it is sufficient if we have
	\[\frac{s}{n}\cdot \frac{\binom{n}{t}}{\binom{k}{t}}\cdot \frac{\binom{r}{t}}{\binom{s}{t}} \geq 1.\]
	
	Note that 
	\begin{align*}
		\frac{s}{n}\cdot\frac{\binom{n}{t}}{\binom{k}{t}}\cdot \frac{\binom{r}{t}}{\binom{s}{t}} 
		&= \frac{s}{n} \prod_{w = 0}^{t - 1}\frac{(n - w)(r - w)}{(s - w)(k - w)}\\ 
		&= \prod_{w = 1}^{t - 1}\frac{(n - w)(r - w)}{(s - w)(k - w)} \cdot \frac{r}{k} \\
		&> \prod_{w = 1}^{t - 1}\frac{(n - w)(r - w)}{(s - w)(k - w)} \cdot \mu.
	\end{align*}
	Therefore, we only require that
	\[\frac{(n - w)(r - w)}{(s - w)(k - w)} \geq \mu^{-\frac{1}{t - 1}},\quad \forall 1 \leq w \leq t - 1.\]
	
	Let $c := \mu^{-\frac{1}{t - 1}} > 1$, the above condition is equivalent to 
	\[c(s - w)(k - w) - (n - w)(r - w) \leq 0,\quad \forall 1 \leq w \leq t - 1,\]
	and by the property of quadratic functions, we only need to work on the case of $w = 1$ and $w = t - 1$.
	
	When $w = 1$, the condition becomes 
	\[\frac{s - 1}{n - 1} \leq \frac{r - 1}{c(k - 1)}.\]
	Note that $\frac{s - 1}{n - 1} < \frac{s}{n} = \gamma$, hence it is sufficient with
	$\gamma \leq \frac{r - 1}{c(k - 1)}$. When $\mu \geq 1/2$, $r \geq \max\{t, 3\}$ and $n \geq 3k - 3$, we have $c = \mu^{-\frac{1}{r - 1}} < 3/2$. Therefore, 
	\begin{align*}
		s \leq \frac{n}{c\cdot k} + t - 1 &\Longrightarrow \gamma \leq \frac{1}{c\cdot k} + \frac{t - 1}{n} \\
		&\Longrightarrow \gamma \leq \frac{1}{c(k - 1)} + \frac{2(r - 2)}{n} \\
		&\Longrightarrow \gamma \leq \frac{1}{c(k - 1)} + \frac{2(r - 2)}{3(k - 1)} \\
		&\Longrightarrow \gamma \leq \frac{1}{c(k - 1)} + \frac{r - 2}{c(k - 1)} \\
		&\Longrightarrow \gamma \leq \frac{r - 1}{c(k - 1)}.
	\end{align*}
	
	When $w = t - 1$, the condition becomes
	\[(n - t + 1)(r - t + 1) \geq c(s - t + 1)(k - t + 1), \]
	or
	\[s \leq \frac{(n - t + 1)(r - t + 1)}{c(k - t + 1)} + t - 1.\]
	
	which naturally establishes when $s \leq \frac{n}{ck} + t - 1$, as $\frac{n}{k} \leq \frac{n - t + 1}{k - t + 1}$ and $t\leq r$.
	\qed
\end{proof}

%

\subsection{Proof of Theorem~\protect{\ref{thm:random-sampling}}}\label{app:proof-thm-random-sampling}
\begin{proof}
	To show the result, first consider a specific set of corrupted nodes with size precisely $\gamma n$. We denote the set as $P$. Further we use $g(P)$ to denote the number of corrupted groups with nodes $P$ corrupted under the group assignment scheme $\mathcal{A}$. Clearly, $g(P) \leq \maxcorruptedgroupnumber$ by definition.
	
	Now suppose we uniformly draw $\beta m$ groups from $\mathcal{A}$ to obtain $\mathcal{A}'$, and let $X$ be a random variable indicating the number of corrupted groups in the new scheme $\mathcal{A}'$. By hypergeometric tail bound, we have
	\begin{align*}
		&\Pr\left[X \geq \left(\frac{\maxcorruptedgroupnumber}{m} + \delta\right)\cdot \beta m\right] \\
		=\ &\Pr\left[X \geq \left[\frac{g(P)}{m} + \left(\frac{\maxcorruptedgroupnumber - g(P)}{m} + \delta\right)\right] \cdot \beta m \right] \\
		\leq\ &\exp\left(-2\beta m\cdot\left(\frac{\maxcorruptedgroupnumber - g(P)}{m} + \delta \right)^2\right) \\
		\leq\ &\exp(-2\beta m \cdot \delta^2).
	\end{align*}
	Here, $\delta > 0$ is a parameter to be determined. The probability is over all possible choices of $\mathcal{A}'$. Let $\maxcorruptedgroupnumbernew$ denote the maximal number of corrupted groups under a $\gamma$-adversary in the new scheme $\mathcal{A}'$. By a union bound, we have
	\[\Pr\left[f(\gamma; S, \mathcal{A}',\mu) \geq \beta \maxcorruptedgroupnumber + \delta \cdot \beta m\right] \leq \exp(-2\beta m \cdot \delta^2) \binom{n}{\gamma n}.\]
	
	By Stirling's formula (see~\cite{Robbins55}), 
	\begin{align*}
	\binom{n}{\gamma n} &= \frac{n!}{(\gamma n)!(n - \gamma n)!} \\
	&\leq \frac{e}{2\pi} \frac{n^{n + \frac{1}{2}}}{(\gamma n)^{\gamma n + \frac{1}{2}}(n - \gamma n)^{n - \gamma n + \frac{1}{2}}} \\
	&= \frac{e}{2\pi} \sqrt{\frac{1}{n \gamma (1 - \gamma)}}\cdot \left(\frac{1}{\gamma^{\gamma}(1 - \gamma)^{ 1 - \gamma}}\right)^n \\
	&\leq \frac{e}{2\pi} \left(\frac{1}{\gamma^{\gamma}(1 - \gamma)^{ 1 - \gamma}}\right)^n.
	\end{align*}
	
	Hence, 
	\begin{align*}
		\Pr[\maxcorruptedgroupnumbernew \geq \beta \maxcorruptedgroupnumber + \delta \cdot \beta m] &\leq \exp(-2\beta m \cdot \delta^2) \binom{n}{\gamma n} \\
		&\leq \frac{e}{2\pi} \exp(-2\beta m \cdot \delta^2 + n H(\gamma)).
	\end{align*}
	
	Let $\beta m \cdot \delta^2 = \left(\frac{1 + c}{2}\right)\cdot nH(\gamma)$, then w.p. no less than $1 - \frac{e}{2\pi}\exp(-cnH(\gamma))$, we have $\maxcorruptedgroupnumbernew \leq \beta \maxcorruptedgroupnumber + \delta \cdot \beta m$. Under such case, the efficiency factor $\eta'$ of the custody scheme induced by $\mathcal{A}'$ and $\mu$ against a $\gamma$-adversary is lower bounded by
	\begin{align*}
		\eta' &= \gamma \frac{\beta m}{\maxcorruptedgroupnumbernew} - 1 \\
		&\geq \gamma \frac{m}{\maxcorruptedgroupnumber + \delta \cdot m} - 1 \\ &= \gamma \frac{m}{\frac{\gamma\cdot m}{\eta + 1} + \delta \cdot m} - 1 \\
		&= \gamma \frac{\eta + 1}{\gamma + \delta \cdot (\eta + 1)} - 1 \\
		&= \frac{\gamma \sqrt{\beta m}(\eta + 1)}{\gamma \sqrt{\beta m} + \sqrt{(1 + c)n H(\gamma) / 2}(\eta + 1)} - 1.
	\end{align*}
	The third line is due to the definition of $\eta$:
	\[\eta = \gamma \cdot \frac{m}{\maxcorruptedgroupnumber} - 1.\]
	\qed
\end{proof}

\section{Experimental Results}\label{app:experimental-results}

As we have already shown in Theorem~\ref{thm:hardness}, given a group assignment scheme and the number of corrupted nodes $s$, it is generally computationally involved to figure out the optimal subset to corrupt and the efficiency factor for the corresponding custody scheme. In Section~\ref{sec:construction} and Appendix~\ref{app:multi-layer-sharding-design}, we analyze four types of concrete design, and for those three types except symmetric design, we successfully propose lower bounds on the efficiency factor theoretically. In this section, we give an estimation of the performance of these custody schemes from an experimental view. 

\begin{interposition}[Settings]
	Given a custody scheme induced by $\mathcal{A}$ and $\mu$, to estimate the value $\maxcorruptedgroupnumber$ given adversary power $\gamma$, we uniformly pick a subset $P$ consisting of $s = \gamma n$ nodes as the corrupted node set, and calculate the number of corrupted groups. We independently repeat such process for $500,000$ times, and record the maximal number of corrupted groups in all these trials. We deem such value as the estimation of the real value $\maxcorruptedgroupnumber$, and correspondingly compute the efficiency factor $\eta$.
\end{interposition}

Our estimation of $\eta$ is certainly an upper bound on the real value, since the maximal number of corrupted sets we can find never exceeds $\maxcorruptedgroupnumber$. However, we can expect that our estimation is close to the real value due to our large number of attempts. When the number of custodians $n$ is rather small, we can even guarantee that the optimal corrupting strategy is reached with high probability, therefore we come up with the accurate $\eta$ value.

\begin{table}[!t]
	\caption{Estimation on efficiency factor $\eta$ of custody schemes induced by different polynomial designs $\polynomialdesign$ and $\mu \in \{1/2, 2/3\}$ against adversary with power $\gamma \in \{1/2\cdot \mu, 2/3\cdot \mu\}$.}
	\centering
	\begin{tabular}{c*{2}{p{1.5em}<{\centering}}*{1}{p{3em}<{\centering}}*{4}{p{5em}<{\centering}}}	
		\toprule
		\multicolumn{4}{c}{parameters} & \multicolumn{4}{c}{estimation of $\eta$} \\
		\cmidrule(lr){1-4}\cmidrule(lr){5-8}
		\multirow{2}*{$(k, q, d)$} & \multirow{2}*{$k$} & \multirow{2}*{$n$} & \multirow{2}*{$m$} & \multicolumn{2}{c}{$\mu = 1/2$} & \multicolumn{2}{c}{$\mu = 2/3$} \\
		\cmidrule(lr){5-6}\cmidrule(lr){7-8}
		& & & & $\gamma = 1/2\cdot \mu$ & $\gamma = 2/3\cdot \mu$ & $\gamma = 1/2\cdot \mu$ & $\gamma = 2/3\cdot \mu$ \\
		\midrule
		(5, 5, 2) & 5 & 25 & 25 & 0.500000 & 0.142857 & 3.000000 & 1.200000 \\ 
		(5, 5, 3) & 5 & 25 & 125 & 1.307692 & 0.379310 & 5.666667 & 1.894737 \\ 
		(5, 5, 4) & 5 & 25 & 625 & 1.542373 & 0.612903 & 6.142857 & 2.125000 \\ 
		(5, 7, 2) & 5 & 35 & 49 & 0.400000 & 0.100000 & 1.566667 & 1.333333 \\ 
		(5, 7, 3) & 5 & 35 & 343 & 1.240000 & 0.437333 & 5.341176 & 2.062500 \\ 
		(5, 7, 4) & 5 & 35 & 2,401 & 1.744000 & 0.585294 & 5.987037 & 2.166154 \\ 
		(5, 11, 2) & 5 & 55 & 121 & 0.682353 & 0.164706 & 2.600000 & 1.514286 \\ 
		(5, 11, 3) & 5 & 55 & 1,331 & 1.231206 & 0.491781 & 5.405882 & 2.244693 \\ 
		(5, 11, 4) & 5 & 55 & 14,641 & 1.535238 & 0.537741 & 5.729775 & 2.427468 \\ 
		(7, 7, 2) & 7 & 49 & 49 & 1.000000 & 0.230769 & 2.200000 & 0.909091 \\ 
		(7, 7, 3) & 7 & 49 & 343 & 1.709677 & 0.600000 & 4.333333 & 1.672727 \\ 
		(7, 7, 4) & 7 & 49 & 2,401 & 2.418605 & 0.912195 & 5.533333 & 2.099398 \\ 
		(7, 11, 2) & 7 & 77 & 121 & 1.132653 & 0.403061 & 2.928571 & 1.054945 \\ 
		(7, 11, 3) & 7 & 77 & 1,331 & 2.041005 & 0.815726 & 5.449893 & 1.746328 \\ 
		(7, 11, 4) & 7 & 77 & 14,641 & 2.552325 & 0.983961 & 6.521474 & 2.039425 \\ 
		(7, 13, 2) & 7 & 91 & 169 & 1.269841 & 0.428571 & 2.714286 & 1.122449 \\ 
		(7, 13, 3) & 7 & 91 & 2,197 & 2.299024 & 0.801706 & 5.525097 & 1.815494 \\ 
		(11, 11, 2) & 11 & 121 & 121 & 2.333333 & 0.818182 & 7.000000 & 3.076923 \\ 
		(11, 11, 3) & 11 & 121 & 1,331 & 2.000000 & 0.379310 & 11.941176 & 2.491018 \\ 
		(11, 11, 4) & 11 & 121 & 14,641 & 2.175853 & 0.368778 & 13.069767 & 2.658300 \\ 
		(11, 13, 2) & 11 & 143 & 169 & 2.446970 & 0.851515 & 6.935065 & 3.379679 \\ 
		(11, 13, 3) & 11 & 143 & 2,197 & 2.004063 & 0.413094 & 12.886364 & 2.611601 \\ 
		(11, 17, 2) & 11 & 187 & 289 & 2.949495 & 1.038685 & 9.646465 & 3.933566 \\ 
		(11, 17, 3) & 11 & 143 & 4,913 & 2.443149 & 0.462216 & 13.543831 & 2.893994 \\ 
		\bottomrule
	\end{tabular}
	\label{tab:polynomial-design}
\end{table}

\begin{table}[!t]
	\caption{Estimation on efficiency factor $\eta$ of custody schemes induced by different block designs $\blockdesign$ and $\mu \in \{1/2, 2/3\}$ against adversary with power $\gamma \in \{1/2\cdot \mu, 2/3\cdot \mu\}$.}
	\centering
	\begin{tabular}{c*{2}{p{1.5em}<{\centering}}*{1}{p{3em}<{\centering}}*{4}{p{5em}<{\centering}}}	
		\toprule
		\multicolumn{4}{c}{parameters} & \multicolumn{4}{c}{estimation of $\eta$} \\
		\cmidrule(lr){1-4}\cmidrule(lr){5-8}
		\multirow{2}*{$t$-$(n, k, \lambda)$} & \multirow{2}*{$k$} & \multirow{2}*{$n$} & \multirow{2}*{$m$} & \multicolumn{2}{c}{$\mu = 1/2$} & \multicolumn{2}{c}{$\mu = 2/3$} \\
		\cmidrule(lr){5-6}\cmidrule(lr){7-8}
		& & & & $\gamma = 1/2\cdot \mu$ & $\gamma = 2/3\cdot \mu$ & $\gamma = 1/2\cdot \mu$ & $\gamma = 2/3\cdot \mu$ \\
		\midrule
		2-(25, 5, 1) & 5 & 25 & 30 & 0.800000 & 0.371429 & 3.800000 & 1.640000 \\ 
		2-(25, 5, 2) & 5 & 25 & 60 & 0.800000 & 0.371429 & 3.800000 & 1.640000 \\ 
		2-(25, 5, 3) & 5 & 25 & 90 & 0.800000 & 0.371429 & 3.800000 & 1.640000 \\ 
		2-(35, 5, 2) & 5 & 35 & 119 & 0.942857 & 0.289655 & 3.155556 & 1.684211 \\ 
		2-(35, 5, 4) & 5 & 35 & 238 & 1.176000 & 0.411321 & 3.986667 & 2.000000 \\ 
		2-(35, 5, 6) & 5 & 35 & 357 & 1.266667 & 0.476316 & 4.610000 & 2.187500 \\ 
		2-(55, 5, 2) & 5 & 55 & 297 & 1.064706 & 0.350000 & 3.628571 & 1.880000 \\ 
		2-(55, 5, 4) & 5 & 55 & 594 & 1.301639 & 0.418978 & 4.717647 & 2.200000 \\ 
		2-(55, 5, 6) & 5 & 55 & 891 & 1.366292 & 0.487755 & 5.075000 & 2.380870 \\ 
		2-(125, 5, 1) & 5 & 125 & 775 & 1.044681 & 0.412222 & 4.408511 & 2.072072 \\ 
		2-(125, 5, 2) & 5 & 125 & 1,550 & 1.171751 & 0.469364 & 4.981176 & 2.216981 \\ 
		2-(125, 5, 3) & 5 & 125 & 2,325 & 1.209195 & 0.507115 & 5.408403 & 2.332248 \\ 
		2-(49, 7, 1) & 7 & 49 & 56 & 0.959184 & 0.306122 & 2.047619 & 1.000000 \\ 
		2-(49, 7, 2) & 7 & 49 & 112 & 1.285714 & 0.462857 & 3.063492 & 1.400000 \\ 
		2-(49, 7, 3) & 7 & 49 & 168 & 1.571429 & 0.567347 & 3.571429 & 1.571429 \\ 
		2-(77, 7, 3) & 7 & 77 & 418 & 1.714286 & 0.675485 & 4.219780 & 1.528376 \\ 
		2-(77, 7, 6) & 7 & 77 & 836 & 2.033613 & 0.762523 & 4.775076 & 1.674948 \\ 
		2-(77, 7, 9) & 7 & 77 & 1,254 & 2.257143 & 0.833977 & 5.076759 & 1.796537 \\ 
		2-(91, 7, 1) & 7 & 91 & 195 & 1.481203 & 0.495017 & 3.017857 & 1.380952 \\ 
		2-(91, 7, 2) & 7 & 91 & 390 & 1.857143 & 0.607143 & 3.945055 & 1.484472 \\ 
		2-(91, 7, 3) & 7 & 91 & 585 & 2.009119 & 0.634383 & 4.075188 & 1.650957 \\ 
		2-(121, 11, 1) & 11 & 121 & 132 & 2.636364 & 0.897233 & 6.272727 & 3.447552 \\ 
		2-(121, 11, 2) & 11 & 121 & 264 & 3.090909 & 1.128603 & 9.909091 & 4.256198 \\ 
		2-(121, 11, 3) & 11 & 121 & 396 & 3.675325 & 1.218798 & 12.090909 & 4.595308 \\ 
		3-(26, 6, 1) & 6 & 26 & 130 & 9.000000 & 2.333333 & 19.000000 & 6.857143 \\ 
		3-(26, 6, 2) & 6 & 26 & 260 & 9.000000 & 2.333333 & 19.000000 & 6.857143 \\ 
		3-(26, 6, 3) & 6 & 26 & 390 & 9.000000 & 2.750000 & 19.000000 & 6.857143 \\ 
		3-(50, 8, 1) & 8 & 50 & 350 & 7.400000 & 2.733333 & 11.444444 & 3.812500 \\ 
		3-(50, 8, 2) & 8 & 50 & 700 & 8.333333 & 2.929825 & 15.000000 & 4.310345 \\ 
		3-(50, 8, 3) & 8 & 50 & 1,050 & 9.500000 & 3.253165 & 16.684211 & 4.566265 \\ 
		3-(122, 12, 1) & 12 & 122 & 1,342 & 13.347826 & 3.835165 & 43.000000 & 10.880000 \\ 
		3-(122, 12, 2) & 12 & 122 & 2,684 & 15.500000 & 4.146199 & 54.000000 & 12.200000 \\ 
		3-(122, 12, 3) & 12 & 122 & 4,026 & 17.000000 & 4.387755 & 61.857143 & 12.921875 \\ 
		\bottomrule
	\end{tabular}
	\label{tab:block-design}
\end{table}

\begin{table}[!t]
	\caption{Estimation on efficiency factor $\eta$ of custody schemes induced by different random multi-layer sharding designs $\randommultilayershardingdesign$ and $\mu \in \{1/2, 2/3\}$ against adversary with power $\gamma \in \{1/2\cdot \mu, 2/3\cdot \mu\}$.}
	\centering
	\begin{tabular}{c*{2}{p{1.5em}<{\centering}}*{1}{p{3em}<{\centering}}*{4}{p{5em}<{\centering}}}	
		\toprule
		\multicolumn{4}{c}{parameters} & \multicolumn{4}{c}{estimation of $\eta$} \\
		\cmidrule(lr){1-4}\cmidrule(lr){5-8}
		\multirow{2}*{$(n, k, l)$} & \multirow{2}*{$k$} & \multirow{2}*{$n$} & \multirow{2}*{$m$} & \multicolumn{2}{c}{$\mu = 1/2$} & \multicolumn{2}{c}{$\mu = 2/3$} \\
		\cmidrule(lr){5-6}\cmidrule(lr){7-8}
		& & & & $\gamma = 1/2\cdot \mu$ & $\gamma = 2/3\cdot \mu$ & $\gamma = 1/2\cdot \mu$ & $\gamma = 2/3\cdot \mu$ \\
		\midrule
		\multirow{2}*{(60, 5, 30)} & \multirow{2}*{5} &\multirow{2}*{60} &\multirow{2}*{360} & 0.638530 & 0.303921 & 3.228022 & 1.602892 \\ 
		& & & & ±0.059583 & ±0.037254 & ±0.156593 & ±0.086763 \\ 
		\cmidrule(lr){1-8}
		\multirow{2}*{(60, 5, 60)} & \multirow{2}*{5} &\multirow{2}*{60} &\multirow{2}*{720} & 0.856459 & 0.416641 & 4.069170 & 1.902890 \\ 
		& & & & ±0.038278 & ±0.026397 & ±0.112648 & ±0.040505 \\ 
		\cmidrule(lr){1-8}
		\multirow{2}*{(60, 5, 120)} &  \multirow{2}*{5} &\multirow{2}*{60} &\multirow{2}*{1,440} & 1.006833 & 0.515100 & 5.042120 & 2.184005 \\ 
		& & & & ±0.050310 & ±0.015101 & ±0.120042 & ±0.032490 \\ 
		\cmidrule(lr){1-8}
		\multirow{2}*{(120, 5, 60)} & \multirow{2}*{5} &\multirow{2}*{120} &\multirow{2}*{1,440} & 0.961867 & 0.429171 & 4.412032 & 1.965180 \\ 
		& & & & ±0.005346 & ±0.015273 & ±0.093850 & ±0.020736 \\ 
		\cmidrule(lr){1-8}
		\multirow{2}*{(120, 5, 120)} & \multirow{2}*{5} &\multirow{2}*{120} &\multirow{2}*{2,880} & 1.096430 & 0.501281 & 5.022343 & 2.156405 \\ 
		& & & & ±0.027464 & ±0.010835 & ±0.135552 & ±0.015665 \\ 
		\cmidrule(lr){1-8}
		\multirow{2}*{(120, 5, 240)} & \multirow{2}*{5} &\multirow{2}*{120} &\multirow{2}*{5,760} & 1.208610 & 0.547134 & 5.686482 & 2.304035 \\ 
		& & & & ±0.006775 & ±0.006393 & ±0.071641 & ±0.021454 \\ 
		\cmidrule(lr){1-8}	
		\multirow{2}*{(70, 7, 35)} & \multirow{2}*{7} &\multirow{2}*{70} &\multirow{2}*{350} & 0.547501 & 0.204222 & 2.709677 & 1.369865 \\ 
		& & & & ±0.056273 & ±0.006304 & ±0.000000 & ±0.090453 \\ 
		\cmidrule(lr){1-8}
		\multirow{2}*{(70, 7, 70)} & \multirow{2}*{7} &\multirow{2}*{70} &\multirow{2}*{700} & 0.839182 & 0.303377 & 3.601841 & 1.589808 \\ 
		& & & & ±0.049708 & ±0.018461 & ±0.092037 & ±0.129490 \\ 
		\cmidrule(lr){1-8}
		\multirow{2}*{(70, 7, 140)} & \multirow{2}*{7} &\multirow{2}*{70} &\multirow{2}*{1,400} & 1.029869 & 0.394054 & 4.476967 & 1.904762 \\ 
		& & & & ±0.006059 & ±0.012673 & ±0.065202 & ±0.047619 \\ 
		\cmidrule(lr){1-8}
		\multirow{2}*{(140, 7, 70)} & \multirow{2}*{7} &\multirow{2}*{140} &\multirow{2}*{1,400} & 0.872997 & 0.383612 & 4.143678 & 1.870924 \\ 
		& & & & ±0.050080 & ±0.014564 & ±0.143678 & ±0.039874 \\ 
		\cmidrule(lr){1-8}
		\multirow{2}*{(140, 7, 140)} & \multirow{2}*{7} &\multirow{2}*{140} &\multirow{2}*{2,800} & 1.035038 & 0.449983 & 4.699272 & 2.065555 \\ 
		& & & & ±0.017748 & ±0.005713 & ±0.123513 & ±0.011368 \\ 
		\cmidrule(lr){1-8}
		\multirow{2}*{(140, 7, 280)} & \multirow{2}*{7} &\multirow{2}*{140} &\multirow{2}*{5,600} & 1.162266 & 0.507011 & 5.358080 & 2.225164 \\ 
		& & & & ±0.015028 & ±0.008640 & ±0.120792 & ±0.025164 \\ 
		\cmidrule(lr){1-8}
		\multirow{2}*{(110, 11, 55)} & \multirow{2}*{11} &\multirow{2}*{110} &\multirow{2}*{550} & 0.646586 & 0.229027 & 3.090909 & 1.540323 \\ 
		& & & & ±0.020081 & ±0.020973 & ±0.000000 & ±0.040323 \\ 
		\cmidrule(lr){1-8}
		\multirow{2}*{(110, 11, 110)} & \multirow{2}*{11} &\multirow{2}*{110} &\multirow{2}*{1,100} & 0.837075 & 0.333406 & 3.872874 & 1.816436 \\ 
		& & & & ±0.024994 & ±0.009877 & ±0.197549 & ±0.057815 \\ 
		\cmidrule(lr){1-8}
		\multirow{2}*{(110, 11, 220)} & \multirow{2}*{11} &\multirow{2}*{110} &\multirow{2}*{2,200} & 1.012352 & 0.417328 & 4.691789 & 2.014730 \\ 
		& & & & ±0.048716 & ±0.002790 & ±0.022498 & ±0.042595 \\ 
		\bottomrule
	\end{tabular}
	\label{tab:multi-layer-sharding-design}
\end{table}

\begin{table}[!t]
	\caption{Comparison of estimation of efficiency factor $\eta$ of custody schemes induced by different designs with $k = 5$ and $\mu \in \{1/2, 2/3\}$ against adversary with power $\gamma \in \{1/2\cdot \mu, 2/3\cdot \mu\}$.}
	\centering
	\begin{tabular}{*{2}{p{1.5em}<{\centering}}cc*{4}{p{5em}<{\centering}}}	
		\toprule
		\multicolumn{4}{c}{parameters} & \multicolumn{4}{c}{estimation of $\eta$} \\
		\cmidrule(lr){1-4}\cmidrule(lr){5-8}
		\multirow{2}*{$n$} & \multirow{2}*{$k$} & \multirow{2}*{design type} & \multirow{2}*{$m$} & \multicolumn{2}{c}{$\mu = 1/2$} & \multicolumn{2}{c}{$\mu = 2/3$} \\
		\cmidrule(lr){5-6}\cmidrule(lr){7-8}
		& & & & $\gamma = 1/2\cdot \mu$ & $\gamma = 2/3\cdot \mu$ & $\gamma = 1/2\cdot \mu$ & $\gamma = 2/3\cdot \mu$ \\
		\midrule
		\multirow{13}*{25} & \multirow{13}*{5} & sym & 53,130 & 1.476923 & 0.564948 & 7.929412 & 2.528767 \\ 
		\cmidrule(lr){3-8}
		& & \multirow{3}*{poly} & 25 & 0.500000 & 0.142857 & 3.000000 & 1.200000 \\ 
		& & & 125 & 1.307692 & 0.379310 & 5.666667 & 1.894737 \\ 
		& & & 625 & 1.542373 & 0.612903 & 6.142857 & 2.125000 \\ 
		\cmidrule(lr){3-8}
		& & \multirow{3}*{blck} & 30 & 0.800000 & 0.371429 & 3.800000 & 1.640000 \\ 
		& & & 60 & 0.800000 & 0.371429 & 3.800000 & 1.640000 \\ 
		& & & 90 & 0.800000 & 0.371429 & 3.800000 & 1.640000 \\
		\cmidrule(lr){3-8}
		& & \multirow{6}*{rmls} & \multirow{2}*{60} & 0.168831 & -0.037594 & 1.266666 & 0.767857 \\ 
		& & & & ±0.140260 & ±0.048120 & ±0.133333 & ±0.117857 \\ 
		& & & \multirow{2}*{125} & 0.441647 & 0.143791 & 2.333333 & 1.245833 \\
		& & & & ±0.137299 & ±0.032680 & ±0.000000 & ±0.045833 \\ 
		& & & \multirow{2}*{250} & 0.819853 & 0.311828 & 3.458204 & 1.591464 \\
		& & & & ±0.055147 & ±0.021505 & ±0.247678 & ±0.091464 \\ 
		\cmidrule(lr){1-8}
		\multirow{13}*{35} & \multirow{13}*{5} & sym & 324,632 & 1.718937 & 0.640733 & 8.140379 & 2.810291 \\ 
		\cmidrule(lr){3-8}
		& & \multirow{3}*{poly} & 49 & 0.400000 & 0.100000 & 1.566667 & 1.333333 \\ 
		& & & 343 & 1.240000 & 0.437333 & 5.341176 & 2.062500 \\ 
		& & & 2,401 & 1.744000 & 0.585294 & 5.987037 & 2.166154 \\ 
		\cmidrule(lr){3-8}
		& & \multirow{3}*{blck} & 119 & 0.942857 & 0.289655 & 3.155556 & 1.684211 \\ 
		& & & 238 & 1.176000 & 0.411321 & 3.986667 & 2.000000 \\ 
		& & & 357 & 1.266667 & 0.476316 & 4.610000 & 2.187500 \\
		\cmidrule(lr){3-8}
		& & \multirow{6}*{rmls} & \multirow{2}*{85} & 0.395790 & 0.116667 & 1.996795 & 1.082500 \\ 
		& & & & ±0.035790 & ±0.016667 & ±0.119872 & ±0.042500 \\ 
		& & & \multirow{2}*{245} & 0.723485 & 0.263653 & 2.776316 & 1.533430 \\
		& & & & ±0.026515 & ±0.041431 & ±0.276316 & ±0.091570 \\ 
		& & & \multirow{2}*{490} & 1.037037 & 0.400463 & 4.083334 & 1.938356 \\
		& & & & ±0.037037 & ±0.025463 & ±0.416667 & ±0.061644 \\ 
		\cmidrule(lr){1-8}
		\multirow{13}*{55} & \multirow{13}*{5} & sym & $3.48\times 10^6$ & 1.592806 & 0.577863 & 6.950000 & 2.654982 \\ 
		\cmidrule(lr){3-8}
		& & \multirow{3}*{poly} & 121 & 0.682353 & 0.164706 & 2.600000 & 1.514286 \\ 
		& & & 1,331 & 1.231206 & 0.491781 & 5.405882 & 2.244693 \\ 
		& & & 14,641 & 1.535238 & 0.537741 & 5.729775 & 2.427468 \\  
		\cmidrule(lr){3-8}
		& & \multirow{3}*{blck} & 297 & 1.064706 & 0.350000 & 3.628571 & 1.880000 \\ 
		& & & 594 & 1.301639 & 0.418978 & 4.717647 & 2.200000 \\ 
		& & & 891 & 1.366292 & 0.487755 & 5.075000 & 2.380870 \\
		\cmidrule(lr){3-8}
		& & \multirow{6}*{rmls} & \multirow{2}*{297} & 0.712195 & 0.231169 & 2.744000 & 1.448770 \\ 
		& & & & ±0.000000 & ±0.031169 & ±0.144000 & ±0.092406 \\ 
		& & & \multirow{2}*{605} & 0.920241 & 0.347500 & 3.664634 & 1.809782 \\
		& & & & ±0.038662 & ±0.027500 & ±0.164634 & ±0.059782 \\ 
		& & & \multirow{2}*{1,210}& 1.151470 & 0.443011 & 4.540900 & 2.106312 \\
		& & & & ±0.048530 & ±0.023656 & ±0.116242 & ±0.036545 \\ 
		\bottomrule
	\end{tabular}
	\label{tab:comparison-size-5}
\end{table}

\begin{table}[!t]
	\caption{Comparison of estimation of efficiency factor $\eta$ of custody schemes induced by different designs with $k = 7$ and $\mu \in \{1/2, 2/3\}$ against adversary with power $\gamma \in \{1/2\cdot \mu, 2/3\cdot \mu\}$.}
	\centering
	\begin{tabular}{*{2}{p{1.5em}<{\centering}}cc*{4}{p{5em}<{\centering}}}	
		\toprule
		\multicolumn{4}{c}{parameters} & \multicolumn{4}{c}{estimation of $\eta$} \\
		\cmidrule(lr){1-4}\cmidrule(lr){5-8}
		\multirow{2}*{$n$} & \multirow{2}*{$k$} & \multirow{2}*{design type} & \multirow{2}*{$m$} & \multicolumn{2}{c}{$\mu = 1/2$} & \multicolumn{2}{c}{$\mu = 2/3$} \\
		\cmidrule(lr){5-6}\cmidrule(lr){7-8}
		& & & & $\gamma = 1/2\cdot \mu$ & $\gamma = 2/3\cdot \mu$ & $\gamma = 1/2\cdot \mu$ & $\gamma = 2/3\cdot \mu$ \\
		\midrule 
		\multirow{13}*{49} & \multirow{13}*{7} & sym & $8.59\times 10^7$ & 2.811410 & 0.949193 & 7.552244 & 2.374737 \\ 
		\cmidrule(lr){3-8}
		& & \multirow{3}*{poly} & 49 & 1.000000 & 0.230769 & 2.200000 & 0.909091 \\ 
		& & & 343 & 1.709677 & 0.600000 & 4.333333 & 1.672727 \\ 
		& & & 2,401 & 2.418605 & 0.912195 & 5.533333 & 2.099398 \\ 
		\cmidrule(lr){3-8}
		& & \multirow{3}*{blck} & 56 & 0.959184 & 0.306122 & 2.047619 & 1.000000 \\ 
		& & & 112 & 1.285714 & 0.462857 & 3.063492 & 1.400000 \\ 
		& & & 168 & 1.571429 & 0.567347 & 3.571429 & 1.571429 \\
		\cmidrule(lr){3-8}
		& & \multirow{6}*{rmls} & \multirow{2}*{168} & 0.328572 & 0.047273 & 1.984874 & 1.149733 \\ 
		& & & & ±0.042857 & ±0.049870 & ±0.242017 & ±0.032085 \\ 
		& & & \multirow{2}*{343} & 0.573530 & 0.217967 & 2.931034 & 1.537500 \\
		& & & & ±0.073530 & ±0.026478 & ±0.068966 & ±0.087500 \\ 
		& & & \multirow{2}*{686} & 0.847046 & 0.314699 & 3.928433 & 1.913461 \\
		& & & & ±0.040594 & ±0.034699 & ±0.162476 & ±0.086538 \\ 
		\cmidrule(lr){1-8}
		\multirow{13}*{77} & \multirow{13}*{7} & sym & $2.40\times 10^9$ & 2.700478 & 0.992168 & 7.434833 & 2.084306 \\ 
		\cmidrule(lr){3-8}
		& & \multirow{3}*{poly} & 121 & 1.132653 & 0.403061 & 2.928571 & 1.054945 \\ 
		& & & 1,331 & 2.041005 & 0.815726 & 5.449893 & 1.746328 \\ 
		& & & 14,641 & 2.552325 & 0.983961 & 6.521474 & 2.039425 \\ 
		\cmidrule(lr){3-8}
		& & \multirow{3}*{blck} & 418 & 1.714286 & 0.675485 & 4.219780 & 1.528376 \\ 
		& & & 836 & 2.033613 & 0.762523 & 4.775076 & 1.674948 \\ 
		& & & 1,254 & 2.257143 & 0.833977 & 5.076759 & 1.796537 \\
		\cmidrule(lr){3-8}
		& & \multirow{6}*{rmls} & \multirow{2}*{418} & 0.625204 & 0.234684 & 3.005456 & 1.461390 \\ 
		& & & & ±0.038390 & ±0.033673 & ±0.235615 & ±0.032818 \\ 
		& & & \multirow{2}*{847} & 0.850137 & 0.341750 & 3.958577 & 1.781630 \\
		& & & & ±0.032746 & ±0.019636 & ±0.134016 & ±0.051703 \\ 
		& & & \multirow{2}*{1,694} & 1.015874 & 0.443818 & 4.672515 & 1.980837 \\
		& & & & ±0.053433 & ±0.018948 & ±0.116959 & ±0.047504 \\ 
		\cmidrule(lr){1-8}
		\multirow{12}*{91} & \multirow{12}*{7} & sym & $8.09\times 10^9$ & 2.877880 & 0.935919 & 6.933541 & 2.129408 \\ 
		\cmidrule(lr){3-8}
		& & \multirow{2}*{poly} & 169 & 1.269841 & 0.428571 & 2.714286 & 1.122449 \\ 
		& & & 2,197 & 2.299024 & 0.801706 & 5.525097 & 1.815494 \\
		\cmidrule(lr){3-8}
		& & \multirow{3}*{blck} & 195 & 1.481203 & 0.495017 & 3.017857 & 1.380952 \\ 
		& & & 390 & 1.857143 & 0.607143 & 3.945055 & 1.484472 \\ 
		& & & 585 & 2.009119 & 0.634383 & 4.075188 & 1.650957 \\
		\cmidrule(lr){3-8}
		& & \multirow{6}*{rmls} & \multirow{2}*{585} & 0.758573 & 0.273109 & 3.338799 & 1.625000 \\ 
		& & & & ±0.054614 & ±0.012605 & ±0.146252 & ±0.053571 \\ 
		& & & \multirow{2}*{1,183} & 0.972507 & 0.359298 & 4.068354 & 1.873279 \\
		& & & & ±0.013603 & ±0.023681 & ±0.131646 & ±0.031749 \\ 
		& & & \multirow{2}*{2,366} & 1.098645 & 0.435289 & 4.758418 & 2.100188 \\
		& & & & ±0.084561 & ±0.014524 & ±0.106244 & ±0.032342 \\ 
		\bottomrule
	\end{tabular}
	\label{tab:comparison-size-7}
\end{table}

\begin{table}[!t]
	\caption{Comparison of estimation of efficiency factor $\eta$ of custody schemes induced by different designs with $k = 6, 8, 12$ and $\mu \in \{1/2, 2/3\}$ against adversary with power $\gamma \in \{1/2\cdot \mu, 2/3\cdot \mu\}$.}
	\centering
	\begin{tabular}{*{2}{p{1.5em}<{\centering}}cc*{4}{p{5em}<{\centering}}}	
		\toprule
		\multicolumn{4}{c}{parameters} & \multicolumn{4}{c}{estimation of $\eta$} \\
		\cmidrule(lr){1-4}\cmidrule(lr){5-8}
		\multirow{2}*{$n$} & \multirow{2}*{$k$} & \multirow{2}*{design type} & \multirow{2}*{$m$} & \multicolumn{2}{c}{$\mu = 1/2$} & \multicolumn{2}{c}{$\mu = 2/3$} \\
		\cmidrule(lr){5-6}\cmidrule(lr){7-8}
		& & & & $\gamma = 1/2\cdot \mu$ & $\gamma = 2/3\cdot \mu$ & $\gamma = 1/2\cdot \mu$ & $\gamma = 2/3\cdot \mu$ \\
		\midrule 
		\multirow{4}*{26} & \multirow{4}*{6} & sym & 230,230 & 8.698795 & 3.116052 & 34.801887 & 7.846154 \\ 
		\cmidrule(lr){3-8}
		& & \multirow{3}*{blck} & 130 & 9.000000 & 2.333333 & 19.000000 & 6.857143 \\ 
		& & & 260 & 9.000000 & 2.333333 & 19.000000 & 6.857143 \\ 
		& & & 390 & 9.000000 & 2.750000 & 19.000000 & 6.857143 \\
		\cmidrule(lr){1-8}
		\multirow{4}*{50} & \multirow{4}*{8} & sym & $5.37\times 10^8$ & 11.454775 & 3.456257 & 24.350413 & 4.805192 \\ 
		\cmidrule(lr){3-8}  
		& & \multirow{3}*{blck} & 350 & 7.400000 & 2.733333 & 11.444444 & 3.812500 \\ 
		& & & 700 & 8.333333 & 2.929825 & 15.000000 & 4.310345 \\ 
		& & & 1,050 & 9.500000 & 3.253165 & 16.684211 & 4.566265 \\
		\cmidrule(lr){1-8}
		\multirow{4}*{122} & \multirow{4}*{12} & sym & $1.30\times 10^{16}$ & 21.589420 & 4.674237 & 128.107816 & 14.492206 \\ 
		\cmidrule(lr){3-8}
		& & \multirow{3}*{blck} & 1,342 & 13.347826 & 3.835165 & 43.000000 & 10.880000 \\ 
		& & & 2,684 & 15.500000 & 4.146199 & 54.000000 & 12.200000 \\ 
		& & & 4,026 & 17.000000 & 4.387755 & 61.857143 & 12.921875 \\
		\bottomrule
	\end{tabular}
	\label{tab:comparison-size-6-8-12}
\end{table}

\subsection{Polynomial Design}

We first show the result for the polynomial design. Concretely, we consider 24 groups of parameters, in which the $k$ value is confined to $5, 7, 11$, and $q$ and $d$ are limited, for the sake of practice. The estimation values on efficiency factor $\eta$ for the custody scheme induced by these group assignment schemes in conjunction with $\mu = 1/2, 2/3$ under adversary power $\gamma \in \{1/2\cdot \mu, 2/3\cdot \mu\}$ are listed in Table~\ref{tab:polynomial-design}.

In general, we observe that the estimation of $\eta$ decreases with the increase of $\gamma$. Meanwhile, for a custody scheme induced by polynomial design, a larger degree $d$ of polynomials leads to better resistance against a rational adversary when other parameters are fixed. These two results may also suit the real value of $\eta$. 

In Theorem~\ref{thm:polynomial-design}, the lower bound value of $\eta$ is irrelevant with the value of $q$. However, this is not the case in reality. When $k = 7, 11$ and $d\leq 4$, the estimation of $\eta$ increases with a larger value $q$. However, such a phenomenon does not occur with $k = 5$, for which the reason is intriguing.

\subsection{Block Design}

For the block design, we consider $33$ different constructions with $k = 5, 6, 7, 8, 11, 12$, which can all be found in~\cite{colbourn2006handbook}. For two designs with multiple relationships, i.e. with the same value of $t, n, k$ but one with $\lambda$ value a multiple of the other one, the larger design is obtained by repeating the smaller one but with random permutation on all custodians in each copy. The corresponding estimation of $\eta$ for $\mu = 1/2, 2/3$ and adversary power $\gamma \in \{1/2\cdot \mu, 2/3\cdot \mu\}$ are shown in Table~\ref{tab:block-design}.

Similarly, the estimation of $\eta$ decreases with the increase of $\gamma$. A key point is that given a block design, the strategy that repeating the groups with a random permutation on custodians leads to a significant enhancement on the performance of efficiency factor. An easy-to-digest account is that a satisfying corrupting effect on one set of groups may lead to a poor effect on another copy if different corrupting choices vary largely on their effectiveness for each copy.

Further, we remark that in practice, 3-designs (block designs with $t = 3$) behave way better than 2-designs (block designs with $t = 2$) from the aspect of efficiency factor. Such result matches with the discussion we present in Section~\ref{sec:block-design}.

\subsection{Multi-Layer Sharding Design}

At last, we come to the performance of the multi-layer sharding design we discussed in Appendix~\ref{app:multi-layer-sharding-design}. In particular, we continue to focus on random multi-layer sharding designs. Specifically, we consider typical cases for $k = 5, 7, 11$. For each choice of parameter tuple, we uniformly generate 5 concrete group assignment schemes at random and evaluate the efficiency factor. The results are present in Table~\ref{tab:multi-layer-sharding-design}.

Again, the estimation of $\eta$ decreases with the increase of $\gamma$. An interesting finding is that the efficiency factor of the custody scheme induced by random multi-layer sharding design slightly changes with the increase or decrease of the number of custodians $n$ against a fixed adversary power. However, as a supplement to Theorem~\ref{thm:random-multi-layer-sharding-design}, the value of $\eta$ considerably increases when the custody scheme owns more sharding layers or a higher value of $l$. Such behavior is similar to the observation we discussed in the last section for block designs, in that a larger $\lambda$ implies better performance. Correspondingly, our explanation naturally suits this case, if we view the multi-layer sharding design as a special block design with $t = 1$. In brief, a corrupting strategy hardly works well on all several identical group assignment schemes only with the order of custodians randomly permuted.

\subsection{Comparison of Different Designs}

At last, we compare four designs we discussed acccording to the evaluation criteria we proposed in Section~\ref{sec:model}. For a better view, we compare the behavior of custody schemes induced by different designs with identical number of custodians $n$ and group size $k$ in Table~\ref{tab:comparison-size-5}, \ref{tab:comparison-size-7}, \ref{tab:comparison-size-6-8-12}.

We have three major conclusions:
\begin{itemize}
	\item Considering the efficiency factor, symmetric design as a benchmark outperforms the other three designs (polynomial design, block design, and random multi-layer sharding design). In fact, in almost all cases that we study, with identical $n$ and $k$, custody schemes induced by symmetric design and fixed $\mu$ owns a higher efficiency factor comparing with other designs under the same adversary power $\gamma$. However, with a massive number of groups, symmetric design is unacceptable in practice. Nevertheless, the other designs own advantage in the number of groups. Such a result implicates the possible positive correlation between the number of groups and the efficiency factor of a custody scheme. We also wonder that whether the efficiency factor given by symmetric design can be beaten by any other possible designs. Further, an interesting observation is that the difference in efficiency factor between the symmetric design and other three designs diminishes with the increase of adversary power $\gamma$. 
	\item The polynomial design and the block design are comparable in practice, while random multi-layer sharding designs perform worse than these two. However, it remains open that whether some other multi-layer sharding strategies (probably deterministic) may behave better.
	\item Numerically, we achieve some good results. Specifically, with polynomial design and block design, we can obtain an efficiency factor of no less than $5$ with $50$-$100$ custodians and $500$-$2,000$ custodian groups. This result is realizable in practice, and further strongly proves that the idea of a decentralized asset custody scheme indeed owns a bright future.
\end{itemize}
\end{subappendices}

\end{document}